\documentclass[longauth]{aaEC}

\usepackage{graphicx}
\usepackage{natbib}
\usepackage{scalerel}
\usepackage{enumitem}
\usepackage{multirow}
\usepackage{url}
\usepackage{glossaries}
\usepackage{cancel}
\usepackage{float}

\newacronym{dlnn}{DLNN}{Deep Learning Neural Network}
\newacronym{GBDT}{GBDT}{Gradient-Boosted Decision-Trees}
\newacronym{ml}{ML}{machine learning}
\newacronym{ms}{SFMS}{star-forming main-sequence}
\newacronym{sfg}{SFGs}{star-forming galaxies}
\newacronym{sfr}{SFRs}{star-formation rates}
\newacronym{snr}{S/N}{signal-to-noise ratio}
\newacronym{sed}{SED}{spectral energy distribution}
\newacronym{odr}{ODR}{orthogonal distance regression}
\newacronym{pp}{PPs}{physical parameters}
\newacronym{nn}{NNs}{nearest-neighbours}
\newacronym{UV}{UV}{ultraviolet}
\newacronym{NIR}{NIR}{near-infrared}
\newacronym{EDF}{EDFs}{Euclid Deep Fields}
\newacronym{EWS}{EWS}{Euclid Wide Survey}
\newacronym{sSFR}{sSFR}{specific star-formation rate}

\newcommand{\nnpz}{\texttt{nnpz}}

\newcommand{\Mstarwun}{\logten(M_{\ast}/M_\odot)}
\newcommand{\sfrwun}{\logten\,(\mathrm{SFR}/M_\odot\,\mathrm{yr}^{-1})}

\usepackage{txfonts}

\usepackage[pdfencoding=auto,psdextra]{hyperref}
\hypersetup{colorlinks=true, linkcolor=blue, filecolor=magenta, urlcolor=blue, citecolor=blue}
\makeatletter
\renewcommand*\aa@pageof{, page \thepage{} of \pageref*{LastPage}}
\makeatother

\usepackage[utf8]{inputenc}

\usepackage[switch, modulo]{lineno}

\usepackage{euclid}
\usepackage{orcidlink}
\let\orcid\orcidlink

\begin{document}

\title{Euclid Quick Data Release (Q1)}
\subtitle{A first view of the star-forming main sequence in the Euclid Deep Fields}

\author{Euclid Collaboration: A.~Enia\orcid{0000-0002-0200-2857}\thanks{\email{andrea.enia@unibo.it}}\inst{\ref{aff1},\ref{aff2}}
\and L.~Pozzetti\orcid{0000-0001-7085-0412}\inst{\ref{aff2}}
\and M.~Bolzonella\orcid{0000-0003-3278-4607}\inst{\ref{aff2}}
\and L.~Bisigello\orcid{0000-0003-0492-4924}\inst{\ref{aff3}}
\and W.~G.~Hartley\inst{\ref{aff4}}
\and C.~Saulder\orcid{0000-0002-0408-5633}\inst{\ref{aff5},\ref{aff6}}
\and E.~Daddi\orcid{0000-0002-3331-9590}\inst{\ref{aff7}}
\and M.~Siudek\orcid{0000-0002-2949-2155}\inst{\ref{aff8},\ref{aff9}}
\and G.~Zamorani\orcid{0000-0002-2318-301X}\inst{\ref{aff2}}
\and P.~Cassata\orcid{0000-0002-6716-4400}\inst{\ref{aff10},\ref{aff3}}
\and F.~Gentile\orcid{0000-0002-8008-9871}\inst{\ref{aff11},\ref{aff2}}
\and L.~Wang\orcid{0000-0002-6736-9158}\inst{\ref{aff12},\ref{aff13}}
\and G.~Rodighiero\orcid{0000-0002-9415-2296}\inst{\ref{aff10},\ref{aff3}}
\and V.~Allevato\orcid{0000-0001-7232-5152}\inst{\ref{aff14}}
\and P.~Corcho-Caballero\orcid{0000-0001-6327-7080}\inst{\ref{aff13}}
\and H.~Dom\'inguez~S\'anchez\orcid{0000-0002-9013-1316}\inst{\ref{aff15}}
\and C.~Tortora\orcid{0000-0001-7958-6531}\inst{\ref{aff14}}
\and M.~Baes\orcid{0000-0002-3930-2757}\inst{\ref{aff16}}
\and Abdurro'uf\orcid{0000-0002-5258-8761}\inst{\ref{aff17}}
\and A.~Nersesian\orcid{0000-0001-6843-409X}\inst{\ref{aff18},\ref{aff16}}
\and L.~Spinoglio\orcid{0000-0001-8840-1551}\inst{\ref{aff19}}
\and J.~Schaye\orcid{0000-0002-0668-5560}\inst{\ref{aff20}}
\and Y.~Ascasibar\orcid{0000-0003-1577-2479}\inst{\ref{aff21},\ref{aff22}}
\and D.~Scott\orcid{0000-0002-6878-9840}\inst{\ref{aff23}}
\and E.~Duran-Camacho\orcid{0000-0002-3153-0536}\inst{\ref{aff24},\ref{aff8}}
\and S.~Quai\orcid{0000-0002-0449-8163}\inst{\ref{aff25},\ref{aff2}}
\and M.~Talia\orcid{0000-0003-4352-2063}\inst{\ref{aff25},\ref{aff2}}
\and Z.~Mao\orcid{0000-0003-4016-845X}\inst{\ref{aff2}}
\and N.~Aghanim\orcid{0000-0002-6688-8992}\inst{\ref{aff26}}
\and B.~Altieri\orcid{0000-0003-3936-0284}\inst{\ref{aff27}}
\and A.~Amara\inst{\ref{aff28}}
\and S.~Andreon\orcid{0000-0002-2041-8784}\inst{\ref{aff29}}
\and N.~Auricchio\orcid{0000-0003-4444-8651}\inst{\ref{aff2}}
\and H.~Aussel\orcid{0000-0002-1371-5705}\inst{\ref{aff7}}
\and C.~Baccigalupi\orcid{0000-0002-8211-1630}\inst{\ref{aff30},\ref{aff31},\ref{aff32},\ref{aff33}}
\and M.~Baldi\orcid{0000-0003-4145-1943}\inst{\ref{aff1},\ref{aff2},\ref{aff34}}
\and A.~Balestra\orcid{0000-0002-6967-261X}\inst{\ref{aff3}}
\and S.~Bardelli\orcid{0000-0002-8900-0298}\inst{\ref{aff2}}
\and A.~Basset\inst{\ref{aff35}}
\and P.~Battaglia\orcid{0000-0002-7337-5909}\inst{\ref{aff2}}
\and R.~Bender\orcid{0000-0001-7179-0626}\inst{\ref{aff5},\ref{aff6}}
\and A.~Biviano\orcid{0000-0002-0857-0732}\inst{\ref{aff31},\ref{aff30}}
\and A.~Bonchi\orcid{0000-0002-2667-5482}\inst{\ref{aff36}}
\and E.~Branchini\orcid{0000-0002-0808-6908}\inst{\ref{aff37},\ref{aff38},\ref{aff29}}
\and M.~Brescia\orcid{0000-0001-9506-5680}\inst{\ref{aff39},\ref{aff14}}
\and J.~Brinchmann\orcid{0000-0003-4359-8797}\inst{\ref{aff40},\ref{aff41}}
\and S.~Camera\orcid{0000-0003-3399-3574}\inst{\ref{aff42},\ref{aff43},\ref{aff44}}
\and G.~Ca\~nas-Herrera\orcid{0000-0003-2796-2149}\inst{\ref{aff45},\ref{aff46},\ref{aff20}}
\and V.~Capobianco\orcid{0000-0002-3309-7692}\inst{\ref{aff44}}
\and C.~Carbone\orcid{0000-0003-0125-3563}\inst{\ref{aff47}}
\and J.~Carretero\orcid{0000-0002-3130-0204}\inst{\ref{aff48},\ref{aff49}}
\and S.~Casas\orcid{0000-0002-4751-5138}\inst{\ref{aff50}}
\and F.~J.~Castander\orcid{0000-0001-7316-4573}\inst{\ref{aff9},\ref{aff51}}
\and M.~Castellano\orcid{0000-0001-9875-8263}\inst{\ref{aff52}}
\and G.~Castignani\orcid{0000-0001-6831-0687}\inst{\ref{aff2}}
\and S.~Cavuoti\orcid{0000-0002-3787-4196}\inst{\ref{aff14},\ref{aff53}}
\and K.~C.~Chambers\orcid{0000-0001-6965-7789}\inst{\ref{aff54}}
\and A.~Cimatti\inst{\ref{aff55}}
\and C.~Colodro-Conde\inst{\ref{aff24}}
\and G.~Congedo\orcid{0000-0003-2508-0046}\inst{\ref{aff56}}
\and C.~J.~Conselice\orcid{0000-0003-1949-7638}\inst{\ref{aff57}}
\and L.~Conversi\orcid{0000-0002-6710-8476}\inst{\ref{aff58},\ref{aff27}}
\and Y.~Copin\orcid{0000-0002-5317-7518}\inst{\ref{aff59}}
\and F.~Courbin\orcid{0000-0003-0758-6510}\inst{\ref{aff60},\ref{aff61}}
\and H.~M.~Courtois\orcid{0000-0003-0509-1776}\inst{\ref{aff62}}
\and M.~Cropper\orcid{0000-0003-4571-9468}\inst{\ref{aff63}}
\and A.~Da~Silva\orcid{0000-0002-6385-1609}\inst{\ref{aff64},\ref{aff65}}
\and H.~Degaudenzi\orcid{0000-0002-5887-6799}\inst{\ref{aff4}}
\and G.~De~Lucia\orcid{0000-0002-6220-9104}\inst{\ref{aff31}}
\and A.~M.~Di~Giorgio\orcid{0000-0002-4767-2360}\inst{\ref{aff19}}
\and C.~Dolding\orcid{0009-0003-7199-6108}\inst{\ref{aff63}}
\and H.~Dole\orcid{0000-0002-9767-3839}\inst{\ref{aff26}}
\and F.~Dubath\orcid{0000-0002-6533-2810}\inst{\ref{aff4}}
\and C.~A.~J.~Duncan\orcid{0009-0003-3573-0791}\inst{\ref{aff57}}
\and X.~Dupac\inst{\ref{aff27}}
\and S.~Dusini\orcid{0000-0002-1128-0664}\inst{\ref{aff66}}
\and A.~Ealet\orcid{0000-0003-3070-014X}\inst{\ref{aff59}}
\and S.~Escoffier\orcid{0000-0002-2847-7498}\inst{\ref{aff67}}
\and M.~Fabricius\orcid{0000-0002-7025-6058}\inst{\ref{aff5},\ref{aff6}}
\and M.~Farina\orcid{0000-0002-3089-7846}\inst{\ref{aff19}}
\and R.~Farinelli\inst{\ref{aff2}}
\and F.~Faustini\orcid{0000-0001-6274-5145}\inst{\ref{aff52},\ref{aff36}}
\and S.~Ferriol\inst{\ref{aff59}}
\and F.~Finelli\orcid{0000-0002-6694-3269}\inst{\ref{aff2},\ref{aff68}}
\and S.~Fotopoulou\orcid{0000-0002-9686-254X}\inst{\ref{aff69}}
\and M.~Frailis\orcid{0000-0002-7400-2135}\inst{\ref{aff31}}
\and E.~Franceschi\orcid{0000-0002-0585-6591}\inst{\ref{aff2}}
\and P.~Franzetti\inst{\ref{aff47}}
\and S.~Galeotta\orcid{0000-0002-3748-5115}\inst{\ref{aff31}}
\and K.~George\orcid{0000-0002-1734-8455}\inst{\ref{aff6}}
\and B.~Gillis\orcid{0000-0002-4478-1270}\inst{\ref{aff56}}
\and C.~Giocoli\orcid{0000-0002-9590-7961}\inst{\ref{aff2},\ref{aff34}}
\and P.~G\'omez-Alvarez\orcid{0000-0002-8594-5358}\inst{\ref{aff70},\ref{aff27}}
\and J.~Gracia-Carpio\inst{\ref{aff5}}
\and B.~R.~Granett\orcid{0000-0003-2694-9284}\inst{\ref{aff29}}
\and A.~Grazian\orcid{0000-0002-5688-0663}\inst{\ref{aff3}}
\and F.~Grupp\inst{\ref{aff5},\ref{aff6}}
\and L.~Guzzo\orcid{0000-0001-8264-5192}\inst{\ref{aff71},\ref{aff29},\ref{aff72}}
\and S.~Gwyn\orcid{0000-0001-8221-8406}\inst{\ref{aff73}}
\and S.~V.~H.~Haugan\orcid{0000-0001-9648-7260}\inst{\ref{aff74}}
\and J.~Hoar\inst{\ref{aff27}}
\and W.~Holmes\inst{\ref{aff75}}
\and I.~M.~Hook\orcid{0000-0002-2960-978X}\inst{\ref{aff76}}
\and F.~Hormuth\inst{\ref{aff77}}
\and A.~Hornstrup\orcid{0000-0002-3363-0936}\inst{\ref{aff78},\ref{aff79}}
\and P.~Hudelot\inst{\ref{aff80}}
\and K.~Jahnke\orcid{0000-0003-3804-2137}\inst{\ref{aff81}}
\and M.~Jhabvala\inst{\ref{aff82}}
\and B.~Joachimi\orcid{0000-0001-7494-1303}\inst{\ref{aff83}}
\and E.~Keih\"anen\orcid{0000-0003-1804-7715}\inst{\ref{aff84}}
\and S.~Kermiche\orcid{0000-0002-0302-5735}\inst{\ref{aff67}}
\and A.~Kiessling\orcid{0000-0002-2590-1273}\inst{\ref{aff75}}
\and B.~Kubik\orcid{0009-0006-5823-4880}\inst{\ref{aff59}}
\and M.~K\"ummel\orcid{0000-0003-2791-2117}\inst{\ref{aff6}}
\and M.~Kunz\orcid{0000-0002-3052-7394}\inst{\ref{aff85}}
\and H.~Kurki-Suonio\orcid{0000-0002-4618-3063}\inst{\ref{aff86},\ref{aff87}}
\and Q.~Le~Boulc'h\inst{\ref{aff88}}
\and A.~M.~C.~Le~Brun\orcid{0000-0002-0936-4594}\inst{\ref{aff89}}
\and D.~Le~Mignant\orcid{0000-0002-5339-5515}\inst{\ref{aff90}}
\and S.~Ligori\orcid{0000-0003-4172-4606}\inst{\ref{aff44}}
\and P.~B.~Lilje\orcid{0000-0003-4324-7794}\inst{\ref{aff74}}
\and V.~Lindholm\orcid{0000-0003-2317-5471}\inst{\ref{aff86},\ref{aff87}}
\and I.~Lloro\orcid{0000-0001-5966-1434}\inst{\ref{aff91}}
\and G.~Mainetti\orcid{0000-0003-2384-2377}\inst{\ref{aff88}}
\and D.~Maino\inst{\ref{aff71},\ref{aff47},\ref{aff72}}
\and E.~Maiorano\orcid{0000-0003-2593-4355}\inst{\ref{aff2}}
\and O.~Mansutti\orcid{0000-0001-5758-4658}\inst{\ref{aff31}}
\and O.~Marggraf\orcid{0000-0001-7242-3852}\inst{\ref{aff92}}
\and M.~Martinelli\orcid{0000-0002-6943-7732}\inst{\ref{aff52},\ref{aff93}}
\and N.~Martinet\orcid{0000-0003-2786-7790}\inst{\ref{aff90}}
\and F.~Marulli\orcid{0000-0002-8850-0303}\inst{\ref{aff25},\ref{aff2},\ref{aff34}}
\and R.~Massey\orcid{0000-0002-6085-3780}\inst{\ref{aff94}}
\and D.~C.~Masters\orcid{0000-0001-5382-6138}\inst{\ref{aff95}}
\and S.~Maurogordato\inst{\ref{aff96}}
\and E.~Medinaceli\orcid{0000-0002-4040-7783}\inst{\ref{aff2}}
\and S.~Mei\orcid{0000-0002-2849-559X}\inst{\ref{aff97},\ref{aff98}}
\and M.~Melchior\inst{\ref{aff99}}
\and Y.~Mellier\inst{\ref{aff100},\ref{aff80}}
\and M.~Meneghetti\orcid{0000-0003-1225-7084}\inst{\ref{aff2},\ref{aff34}}
\and E.~Merlin\orcid{0000-0001-6870-8900}\inst{\ref{aff52}}
\and G.~Meylan\inst{\ref{aff101}}
\and A.~Mora\orcid{0000-0002-1922-8529}\inst{\ref{aff102}}
\and M.~Moresco\orcid{0000-0002-7616-7136}\inst{\ref{aff25},\ref{aff2}}
\and L.~Moscardini\orcid{0000-0002-3473-6716}\inst{\ref{aff25},\ref{aff2},\ref{aff34}}
\and R.~Nakajima\orcid{0009-0009-1213-7040}\inst{\ref{aff92}}
\and C.~Neissner\orcid{0000-0001-8524-4968}\inst{\ref{aff103},\ref{aff49}}
\and S.-M.~Niemi\inst{\ref{aff45}}
\and J.~W.~Nightingale\orcid{0000-0002-8987-7401}\inst{\ref{aff104}}
\and C.~Padilla\orcid{0000-0001-7951-0166}\inst{\ref{aff103}}
\and S.~Paltani\orcid{0000-0002-8108-9179}\inst{\ref{aff4}}
\and F.~Pasian\orcid{0000-0002-4869-3227}\inst{\ref{aff31}}
\and K.~Pedersen\inst{\ref{aff105}}
\and W.~J.~Percival\orcid{0000-0002-0644-5727}\inst{\ref{aff106},\ref{aff107},\ref{aff108}}
\and V.~Pettorino\inst{\ref{aff45}}
\and S.~Pires\orcid{0000-0002-0249-2104}\inst{\ref{aff7}}
\and G.~Polenta\orcid{0000-0003-4067-9196}\inst{\ref{aff36}}
\and M.~Poncet\inst{\ref{aff35}}
\and L.~A.~Popa\inst{\ref{aff109}}
\and F.~Raison\orcid{0000-0002-7819-6918}\inst{\ref{aff5}}
\and R.~Rebolo\inst{\ref{aff110},\ref{aff111},\ref{aff24}}
\and A.~Renzi\orcid{0000-0001-9856-1970}\inst{\ref{aff10},\ref{aff66}}
\and J.~Rhodes\orcid{0000-0002-4485-8549}\inst{\ref{aff75}}
\and G.~Riccio\inst{\ref{aff14}}
\and E.~Romelli\orcid{0000-0003-3069-9222}\inst{\ref{aff31}}
\and M.~Roncarelli\orcid{0000-0001-9587-7822}\inst{\ref{aff2}}
\and E.~Rossetti\orcid{0000-0003-0238-4047}\inst{\ref{aff1}}
\and B.~Rusholme\orcid{0000-0001-7648-4142}\inst{\ref{aff112}}
\and R.~Saglia\orcid{0000-0003-0378-7032}\inst{\ref{aff6},\ref{aff5}}
\and Z.~Sakr\orcid{0000-0002-4823-3757}\inst{\ref{aff113},\ref{aff114},\ref{aff115}}
\and A.~G.~S\'anchez\orcid{0000-0003-1198-831X}\inst{\ref{aff5}}
\and D.~Sapone\orcid{0000-0001-7089-4503}\inst{\ref{aff116}}
\and B.~Sartoris\orcid{0000-0003-1337-5269}\inst{\ref{aff6},\ref{aff31}}
\and J.~A.~Schewtschenko\orcid{0000-0002-4913-6393}\inst{\ref{aff56}}
\and M.~Schirmer\orcid{0000-0003-2568-9994}\inst{\ref{aff81}}
\and P.~Schneider\orcid{0000-0001-8561-2679}\inst{\ref{aff92}}
\and T.~Schrabback\orcid{0000-0002-6987-7834}\inst{\ref{aff117}}
\and M.~Scodeggio\inst{\ref{aff47}}
\and A.~Secroun\orcid{0000-0003-0505-3710}\inst{\ref{aff67}}
\and G.~Seidel\orcid{0000-0003-2907-353X}\inst{\ref{aff81}}
\and S.~Serrano\orcid{0000-0002-0211-2861}\inst{\ref{aff51},\ref{aff118},\ref{aff9}}
\and P.~Simon\inst{\ref{aff92}}
\and C.~Sirignano\orcid{0000-0002-0995-7146}\inst{\ref{aff10},\ref{aff66}}
\and G.~Sirri\orcid{0000-0003-2626-2853}\inst{\ref{aff34}}
\and J.~Skottfelt\orcid{0000-0003-1310-8283}\inst{\ref{aff119}}
\and L.~Stanco\orcid{0000-0002-9706-5104}\inst{\ref{aff66}}
\and J.~Steinwagner\orcid{0000-0001-7443-1047}\inst{\ref{aff5}}
\and C.~Surace\orcid{0000-0003-2592-0113}\inst{\ref{aff90}}
\and P.~Tallada-Cresp\'{i}\orcid{0000-0002-1336-8328}\inst{\ref{aff48},\ref{aff49}}
\and A.~N.~Taylor\inst{\ref{aff56}}
\and H.~I.~Teplitz\orcid{0000-0002-7064-5424}\inst{\ref{aff95}}
\and I.~Tereno\inst{\ref{aff64},\ref{aff120}}
\and S.~Toft\orcid{0000-0003-3631-7176}\inst{\ref{aff121},\ref{aff122}}
\and R.~Toledo-Moreo\orcid{0000-0002-2997-4859}\inst{\ref{aff123}}
\and F.~Torradeflot\orcid{0000-0003-1160-1517}\inst{\ref{aff49},\ref{aff48}}
\and I.~Tutusaus\orcid{0000-0002-3199-0399}\inst{\ref{aff114}}
\and L.~Valenziano\orcid{0000-0002-1170-0104}\inst{\ref{aff2},\ref{aff68}}
\and J.~Valiviita\orcid{0000-0001-6225-3693}\inst{\ref{aff86},\ref{aff87}}
\and T.~Vassallo\orcid{0000-0001-6512-6358}\inst{\ref{aff6},\ref{aff31}}
\and G.~Verdoes~Kleijn\orcid{0000-0001-5803-2580}\inst{\ref{aff13}}
\and A.~Veropalumbo\orcid{0000-0003-2387-1194}\inst{\ref{aff29},\ref{aff38},\ref{aff37}}
\and Y.~Wang\orcid{0000-0002-4749-2984}\inst{\ref{aff95}}
\and J.~Weller\orcid{0000-0002-8282-2010}\inst{\ref{aff6},\ref{aff5}}
\and A.~Zacchei\orcid{0000-0003-0396-1192}\inst{\ref{aff31},\ref{aff30}}
\and F.~M.~Zerbi\inst{\ref{aff29}}
\and I.~A.~Zinchenko\orcid{0000-0002-2944-2449}\inst{\ref{aff6}}
\and E.~Zucca\orcid{0000-0002-5845-8132}\inst{\ref{aff2}}
\and M.~Ballardini\orcid{0000-0003-4481-3559}\inst{\ref{aff124},\ref{aff125},\ref{aff2}}
\and E.~Bozzo\orcid{0000-0002-8201-1525}\inst{\ref{aff4}}
\and C.~Burigana\orcid{0000-0002-3005-5796}\inst{\ref{aff126},\ref{aff68}}
\and R.~Cabanac\orcid{0000-0001-6679-2600}\inst{\ref{aff114}}
\and A.~Cappi\inst{\ref{aff2},\ref{aff96}}
\and D.~Di~Ferdinando\inst{\ref{aff34}}
\and J.~A.~Escartin~Vigo\inst{\ref{aff5}}
\and L.~Gabarra\orcid{0000-0002-8486-8856}\inst{\ref{aff127}}
\and M.~Huertas-Company\orcid{0000-0002-1416-8483}\inst{\ref{aff24},\ref{aff8},\ref{aff128},\ref{aff129}}
\and J.~Mart\'{i}n-Fleitas\orcid{0000-0002-8594-569X}\inst{\ref{aff102}}
\and S.~Matthew\orcid{0000-0001-8448-1697}\inst{\ref{aff56}}
\and N.~Mauri\orcid{0000-0001-8196-1548}\inst{\ref{aff55},\ref{aff34}}
\and R.~B.~Metcalf\orcid{0000-0003-3167-2574}\inst{\ref{aff25},\ref{aff2}}
\and A.~Pezzotta\orcid{0000-0003-0726-2268}\inst{\ref{aff5}}
\and M.~P\"ontinen\orcid{0000-0001-5442-2530}\inst{\ref{aff86}}
\and C.~Porciani\orcid{0000-0002-7797-2508}\inst{\ref{aff92}}
\and I.~Risso\orcid{0000-0003-2525-7761}\inst{\ref{aff130}}
\and V.~Scottez\inst{\ref{aff100},\ref{aff131}}
\and M.~Sereno\orcid{0000-0003-0302-0325}\inst{\ref{aff2},\ref{aff34}}
\and M.~Tenti\orcid{0000-0002-4254-5901}\inst{\ref{aff34}}
\and M.~Viel\orcid{0000-0002-2642-5707}\inst{\ref{aff30},\ref{aff31},\ref{aff33},\ref{aff32},\ref{aff132}}
\and M.~Wiesmann\orcid{0009-0000-8199-5860}\inst{\ref{aff74}}
\and Y.~Akrami\orcid{0000-0002-2407-7956}\inst{\ref{aff133},\ref{aff134}}
\and I.~T.~Andika\orcid{0000-0001-6102-9526}\inst{\ref{aff135},\ref{aff136}}
\and S.~Anselmi\orcid{0000-0002-3579-9583}\inst{\ref{aff66},\ref{aff10},\ref{aff137}}
\and M.~Archidiacono\orcid{0000-0003-4952-9012}\inst{\ref{aff71},\ref{aff72}}
\and F.~Atrio-Barandela\orcid{0000-0002-2130-2513}\inst{\ref{aff138}}
\and C.~Benoist\inst{\ref{aff96}}
\and K.~Benson\inst{\ref{aff63}}
\and P.~Bergamini\orcid{0000-0003-1383-9414}\inst{\ref{aff71},\ref{aff2}}
\and D.~Bertacca\orcid{0000-0002-2490-7139}\inst{\ref{aff10},\ref{aff3},\ref{aff66}}
\and M.~Bethermin\orcid{0000-0002-3915-2015}\inst{\ref{aff139}}
\and A.~Blanchard\orcid{0000-0001-8555-9003}\inst{\ref{aff114}}
\and L.~Blot\orcid{0000-0002-9622-7167}\inst{\ref{aff140},\ref{aff137}}
\and H.~B\"ohringer\orcid{0000-0001-8241-4204}\inst{\ref{aff5},\ref{aff141},\ref{aff142}}
\and S.~Borgani\orcid{0000-0001-6151-6439}\inst{\ref{aff143},\ref{aff30},\ref{aff31},\ref{aff32},\ref{aff132}}
\and A.~S.~Borlaff\orcid{0000-0003-3249-4431}\inst{\ref{aff144},\ref{aff145}}
\and M.~L.~Brown\orcid{0000-0002-0370-8077}\inst{\ref{aff57}}
\and S.~Bruton\orcid{0000-0002-6503-5218}\inst{\ref{aff146}}
\and A.~Calabro\orcid{0000-0003-2536-1614}\inst{\ref{aff52}}
\and B.~Camacho~Quevedo\orcid{0000-0002-8789-4232}\inst{\ref{aff51},\ref{aff9}}
\and F.~Caro\inst{\ref{aff52}}
\and C.~S.~Carvalho\inst{\ref{aff120}}
\and T.~Castro\orcid{0000-0002-6292-3228}\inst{\ref{aff31},\ref{aff32},\ref{aff30},\ref{aff132}}
\and F.~Cogato\orcid{0000-0003-4632-6113}\inst{\ref{aff25},\ref{aff2}}
\and T.~Contini\orcid{0000-0003-0275-938X}\inst{\ref{aff114}}
\and A.~R.~Cooray\orcid{0000-0002-3892-0190}\inst{\ref{aff147}}
\and O.~Cucciati\orcid{0000-0002-9336-7551}\inst{\ref{aff2}}
\and S.~Davini\orcid{0000-0003-3269-1718}\inst{\ref{aff38}}
\and F.~De~Paolis\orcid{0000-0001-6460-7563}\inst{\ref{aff148},\ref{aff149},\ref{aff150}}
\and G.~Desprez\orcid{0000-0001-8325-1742}\inst{\ref{aff13}}
\and A.~D\'iaz-S\'anchez\orcid{0000-0003-0748-4768}\inst{\ref{aff151}}
\and J.~J.~Diaz\inst{\ref{aff24}}
\and S.~Di~Domizio\orcid{0000-0003-2863-5895}\inst{\ref{aff37},\ref{aff38}}
\and J.~M.~Diego\orcid{0000-0001-9065-3926}\inst{\ref{aff15}}
\and P.-A.~Duc\orcid{0000-0003-3343-6284}\inst{\ref{aff139}}
\and Y.~Fang\inst{\ref{aff6}}
\and A.~G.~Ferrari\orcid{0009-0005-5266-4110}\inst{\ref{aff34}}
\and P.~G.~Ferreira\orcid{0000-0002-3021-2851}\inst{\ref{aff127}}
\and A.~Finoguenov\orcid{0000-0002-4606-5403}\inst{\ref{aff86}}
\and A.~Fontana\orcid{0000-0003-3820-2823}\inst{\ref{aff52}}
\and F.~Fontanot\orcid{0000-0003-4744-0188}\inst{\ref{aff31},\ref{aff30}}
\and A.~Franco\orcid{0000-0002-4761-366X}\inst{\ref{aff149},\ref{aff148},\ref{aff150}}
\and K.~Ganga\orcid{0000-0001-8159-8208}\inst{\ref{aff97}}
\and J.~Garc\'ia-Bellido\orcid{0000-0002-9370-8360}\inst{\ref{aff133}}
\and T.~Gasparetto\orcid{0000-0002-7913-4866}\inst{\ref{aff31}}
\and V.~Gautard\inst{\ref{aff11}}
\and E.~Gaztanaga\orcid{0000-0001-9632-0815}\inst{\ref{aff9},\ref{aff51},\ref{aff152}}
\and F.~Giacomini\orcid{0000-0002-3129-2814}\inst{\ref{aff34}}
\and F.~Gianotti\orcid{0000-0003-4666-119X}\inst{\ref{aff2}}
\and G.~Gozaliasl\orcid{0000-0002-0236-919X}\inst{\ref{aff153},\ref{aff86}}
\and M.~Guidi\orcid{0000-0001-9408-1101}\inst{\ref{aff1},\ref{aff2}}
\and C.~M.~Gutierrez\orcid{0000-0001-7854-783X}\inst{\ref{aff154}}
\and A.~Hall\orcid{0000-0002-3139-8651}\inst{\ref{aff56}}
\and S.~Hemmati\orcid{0000-0003-2226-5395}\inst{\ref{aff112}}
\and C.~Hern\'andez-Monteagudo\orcid{0000-0001-5471-9166}\inst{\ref{aff111},\ref{aff24}}
\and H.~Hildebrandt\orcid{0000-0002-9814-3338}\inst{\ref{aff155}}
\and J.~Hjorth\orcid{0000-0002-4571-2306}\inst{\ref{aff105}}
\and J.~J.~E.~Kajava\orcid{0000-0002-3010-8333}\inst{\ref{aff156},\ref{aff157}}
\and Y.~Kang\orcid{0009-0000-8588-7250}\inst{\ref{aff4}}
\and V.~Kansal\orcid{0000-0002-4008-6078}\inst{\ref{aff158},\ref{aff159}}
\and D.~Karagiannis\orcid{0000-0002-4927-0816}\inst{\ref{aff124},\ref{aff160}}
\and K.~Kiiveri\inst{\ref{aff84}}
\and C.~C.~Kirkpatrick\inst{\ref{aff84}}
\and S.~Kruk\orcid{0000-0001-8010-8879}\inst{\ref{aff27}}
\and J.~Le~Graet\orcid{0000-0001-6523-7971}\inst{\ref{aff67}}
\and L.~Legrand\orcid{0000-0003-0610-5252}\inst{\ref{aff161},\ref{aff162}}
\and M.~Lembo\orcid{0000-0002-5271-5070}\inst{\ref{aff124},\ref{aff125}}
\and F.~Lepori\orcid{0009-0000-5061-7138}\inst{\ref{aff163}}
\and G.~Leroy\orcid{0009-0004-2523-4425}\inst{\ref{aff164},\ref{aff94}}
\and G.~F.~Lesci\orcid{0000-0002-4607-2830}\inst{\ref{aff25},\ref{aff2}}
\and J.~Lesgourgues\orcid{0000-0001-7627-353X}\inst{\ref{aff50}}
\and L.~Leuzzi\orcid{0009-0006-4479-7017}\inst{\ref{aff25},\ref{aff2}}
\and T.~I.~Liaudat\orcid{0000-0002-9104-314X}\inst{\ref{aff165}}
\and A.~Loureiro\orcid{0000-0002-4371-0876}\inst{\ref{aff166},\ref{aff167}}
\and J.~Macias-Perez\orcid{0000-0002-5385-2763}\inst{\ref{aff168}}
\and G.~Maggio\orcid{0000-0003-4020-4836}\inst{\ref{aff31}}
\and M.~Magliocchetti\orcid{0000-0001-9158-4838}\inst{\ref{aff19}}
\and E.~A.~Magnier\orcid{0000-0002-7965-2815}\inst{\ref{aff54}}
\and C.~Mancini\orcid{0000-0002-4297-0561}\inst{\ref{aff47}}
\and F.~Mannucci\orcid{0000-0002-4803-2381}\inst{\ref{aff169}}
\and R.~Maoli\orcid{0000-0002-6065-3025}\inst{\ref{aff170},\ref{aff52}}
\and C.~J.~A.~P.~Martins\orcid{0000-0002-4886-9261}\inst{\ref{aff171},\ref{aff40}}
\and L.~Maurin\orcid{0000-0002-8406-0857}\inst{\ref{aff26}}
\and M.~Miluzio\inst{\ref{aff27},\ref{aff172}}
\and P.~Monaco\orcid{0000-0003-2083-7564}\inst{\ref{aff143},\ref{aff31},\ref{aff32},\ref{aff30}}
\and C.~Moretti\orcid{0000-0003-3314-8936}\inst{\ref{aff33},\ref{aff132},\ref{aff31},\ref{aff30},\ref{aff32}}
\and G.~Morgante\inst{\ref{aff2}}
\and C.~Murray\inst{\ref{aff97}}
\and K.~Naidoo\orcid{0000-0002-9182-1802}\inst{\ref{aff152}}
\and A.~Navarro-Alsina\orcid{0000-0002-3173-2592}\inst{\ref{aff92}}
\and S.~Nesseris\orcid{0000-0002-0567-0324}\inst{\ref{aff133}}
\and F.~Passalacqua\orcid{0000-0002-8606-4093}\inst{\ref{aff10},\ref{aff66}}
\and K.~Paterson\orcid{0000-0001-8340-3486}\inst{\ref{aff81}}
\and L.~Patrizii\inst{\ref{aff34}}
\and A.~Pisani\orcid{0000-0002-6146-4437}\inst{\ref{aff67},\ref{aff173}}
\and D.~Potter\orcid{0000-0002-0757-5195}\inst{\ref{aff163}}
\and M.~Radovich\orcid{0000-0002-3585-866X}\inst{\ref{aff3}}
\and P.-F.~Rocci\inst{\ref{aff26}}
\and S.~Sacquegna\orcid{0000-0002-8433-6630}\inst{\ref{aff148},\ref{aff149},\ref{aff150}}
\and M.~Sahl\'en\orcid{0000-0003-0973-4804}\inst{\ref{aff174}}
\and D.~B.~Sanders\orcid{0000-0002-1233-9998}\inst{\ref{aff54}}
\and E.~Sarpa\orcid{0000-0002-1256-655X}\inst{\ref{aff33},\ref{aff132},\ref{aff32}}
\and C.~Scarlata\orcid{0000-0002-9136-8876}\inst{\ref{aff175}}
\and A.~Schneider\orcid{0000-0001-7055-8104}\inst{\ref{aff163}}
\and M.~Schultheis\inst{\ref{aff96}}
\and D.~Sciotti\orcid{0009-0008-4519-2620}\inst{\ref{aff52},\ref{aff93}}
\and E.~Sellentin\inst{\ref{aff176},\ref{aff20}}
\and F.~Shankar\orcid{0000-0001-8973-5051}\inst{\ref{aff177}}
\and L.~C.~Smith\orcid{0000-0002-3259-2771}\inst{\ref{aff178}}
\and S.~A.~Stanford\orcid{0000-0003-0122-0841}\inst{\ref{aff179}}
\and K.~Tanidis\orcid{0000-0001-9843-5130}\inst{\ref{aff127}}
\and G.~Testera\inst{\ref{aff38}}
\and R.~Teyssier\orcid{0000-0001-7689-0933}\inst{\ref{aff173}}
\and S.~Tosi\orcid{0000-0002-7275-9193}\inst{\ref{aff37},\ref{aff130}}
\and A.~Troja\orcid{0000-0003-0239-4595}\inst{\ref{aff10},\ref{aff66}}
\and M.~Tucci\inst{\ref{aff4}}
\and C.~Valieri\inst{\ref{aff34}}
\and A.~Venhola\orcid{0000-0001-6071-4564}\inst{\ref{aff180}}
\and D.~Vergani\orcid{0000-0003-0898-2216}\inst{\ref{aff2}}
\and G.~Verza\orcid{0000-0002-1886-8348}\inst{\ref{aff181}}
\and P.~Vielzeuf\orcid{0000-0003-2035-9339}\inst{\ref{aff67}}
\and N.~A.~Walton\orcid{0000-0003-3983-8778}\inst{\ref{aff178}}
\and E.~Soubrie\orcid{0000-0001-9295-1863}\inst{\ref{aff26}}}
										   
\institute{Dipartimento di Fisica e Astronomia, Universit\`a di Bologna, Via Gobetti 93/2, 40129 Bologna, Italy\label{aff1}
\and
INAF-Osservatorio di Astrofisica e Scienza dello Spazio di Bologna, Via Piero Gobetti 93/3, 40129 Bologna, Italy\label{aff2}
\and
INAF-Osservatorio Astronomico di Padova, Via dell'Osservatorio 5, 35122 Padova, Italy\label{aff3}
\and
Department of Astronomy, University of Geneva, ch. d'Ecogia 16, 1290 Versoix, Switzerland\label{aff4}
\and
Max Planck Institute for Extraterrestrial Physics, Giessenbachstr. 1, 85748 Garching, Germany\label{aff5}
\and
Universit\"ats-Sternwarte M\"unchen, Fakult\"at f\"ur Physik, Ludwig-Maximilians-Universit\"at M\"unchen, Scheinerstrasse 1, 81679 M\"unchen, Germany\label{aff6}
\and
Universit\'e Paris-Saclay, Universit\'e Paris Cit\'e, CEA, CNRS, AIM, 91191, Gif-sur-Yvette, France\label{aff7}
\and
Instituto de Astrof\'isica de Canarias (IAC); Departamento de Astrof\'isica, Universidad de La Laguna (ULL), 38200, La Laguna, Tenerife, Spain\label{aff8}
\and
Institute of Space Sciences (ICE, CSIC), Campus UAB, Carrer de Can Magrans, s/n, 08193 Barcelona, Spain\label{aff9}
\and
Dipartimento di Fisica e Astronomia "G. Galilei", Universit\`a di Padova, Via Marzolo 8, 35131 Padova, Italy\label{aff10}
\and
CEA Saclay, DFR/IRFU, Service d'Astrophysique, Bat. 709, 91191 Gif-sur-Yvette, France\label{aff11}
\and
SRON Netherlands Institute for Space Research, Landleven 12, 9747 AD, Groningen, The Netherlands\label{aff12}
\and
Kapteyn Astronomical Institute, University of Groningen, PO Box 800, 9700 AV Groningen, The Netherlands\label{aff13}
\and
INAF-Osservatorio Astronomico di Capodimonte, Via Moiariello 16, 80131 Napoli, Italy\label{aff14}
\and
Instituto de F\'isica de Cantabria, Edificio Juan Jord\'a, Avenida de los Castros, 39005 Santander, Spain\label{aff15}
\and
Sterrenkundig Observatorium, Universiteit Gent, Krijgslaan 281 S9, 9000 Gent, Belgium\label{aff16}
\and
Johns Hopkins University 3400 North Charles Street Baltimore, MD 21218, USA\label{aff17}
\and
STAR Institute, University of Li{\`e}ge, Quartier Agora, All\'ee du six Ao\^ut 19c, 4000 Li\`ege, Belgium\label{aff18}
\and
INAF-Istituto di Astrofisica e Planetologia Spaziali, via del Fosso del Cavaliere, 100, 00100 Roma, Italy\label{aff19}
\and
Leiden Observatory, Leiden University, Einsteinweg 55, 2333 CC Leiden, The Netherlands\label{aff20}
\and
Departamento de F\'isica Te\'orica, Facultad de Ciencias, Universidad Aut\'onoma de Madrid, 28049 Cantoblanco, Madrid, Spain\label{aff21}
\and
Centro de Investigaci\'{o}n Avanzada en F\'isica Fundamental (CIAFF), Facultad de Ciencias, Universidad Aut\'{o}noma de Madrid, 28049 Madrid, Spain\label{aff22}
\and
Department of Physics and Astronomy, University of British Columbia, Vancouver, BC V6T 1Z1, Canada\label{aff23}
\and
Instituto de Astrof\'{\i}sica de Canarias, V\'{\i}a L\'actea, 38205 La Laguna, Tenerife, Spain\label{aff24}
\and
Dipartimento di Fisica e Astronomia "Augusto Righi" - Alma Mater Studiorum Universit\`a di Bologna, via Piero Gobetti 93/2, 40129 Bologna, Italy\label{aff25}
\and
Universit\'e Paris-Saclay, CNRS, Institut d'astrophysique spatiale, 91405, Orsay, France\label{aff26}
\and
ESAC/ESA, Camino Bajo del Castillo, s/n., Urb. Villafranca del Castillo, 28692 Villanueva de la Ca\~nada, Madrid, Spain\label{aff27}
\and
School of Mathematics and Physics, University of Surrey, Guildford, Surrey, GU2 7XH, UK\label{aff28}
\and
INAF-Osservatorio Astronomico di Brera, Via Brera 28, 20122 Milano, Italy\label{aff29}
\and
IFPU, Institute for Fundamental Physics of the Universe, via Beirut 2, 34151 Trieste, Italy\label{aff30}
\and
INAF-Osservatorio Astronomico di Trieste, Via G. B. Tiepolo 11, 34143 Trieste, Italy\label{aff31}
\and
INFN, Sezione di Trieste, Via Valerio 2, 34127 Trieste TS, Italy\label{aff32}
\and
SISSA, International School for Advanced Studies, Via Bonomea 265, 34136 Trieste TS, Italy\label{aff33}
\and
INFN-Sezione di Bologna, Viale Berti Pichat 6/2, 40127 Bologna, Italy\label{aff34}
\and
Centre National d'Etudes Spatiales -- Centre spatial de Toulouse, 18 avenue Edouard Belin, 31401 Toulouse Cedex 9, France\label{aff35}
\and
Space Science Data Center, Italian Space Agency, via del Politecnico snc, 00133 Roma, Italy\label{aff36}
\and
Dipartimento di Fisica, Universit\`a di Genova, Via Dodecaneso 33, 16146, Genova, Italy\label{aff37}
\and
INFN-Sezione di Genova, Via Dodecaneso 33, 16146, Genova, Italy\label{aff38}
\and
Department of Physics "E. Pancini", University Federico II, Via Cinthia 6, 80126, Napoli, Italy\label{aff39}
\and
Instituto de Astrof\'isica e Ci\^encias do Espa\c{c}o, Universidade do Porto, CAUP, Rua das Estrelas, PT4150-762 Porto, Portugal\label{aff40}
\and
Faculdade de Ci\^encias da Universidade do Porto, Rua do Campo de Alegre, 4150-007 Porto, Portugal\label{aff41}
\and
Dipartimento di Fisica, Universit\`a degli Studi di Torino, Via P. Giuria 1, 10125 Torino, Italy\label{aff42}
\and
INFN-Sezione di Torino, Via P. Giuria 1, 10125 Torino, Italy\label{aff43}
\and
INAF-Osservatorio Astrofisico di Torino, Via Osservatorio 20, 10025 Pino Torinese (TO), Italy\label{aff44}
\and
European Space Agency/ESTEC, Keplerlaan 1, 2201 AZ Noordwijk, The Netherlands\label{aff45}
\and
Institute Lorentz, Leiden University, Niels Bohrweg 2, 2333 CA Leiden, The Netherlands\label{aff46}
\and
INAF-IASF Milano, Via Alfonso Corti 12, 20133 Milano, Italy\label{aff47}
\and
Centro de Investigaciones Energ\'eticas, Medioambientales y Tecnol\'ogicas (CIEMAT), Avenida Complutense 40, 28040 Madrid, Spain\label{aff48}
\and
Port d'Informaci\'{o} Cient\'{i}fica, Campus UAB, C. Albareda s/n, 08193 Bellaterra (Barcelona), Spain\label{aff49}
\and
Institute for Theoretical Particle Physics and Cosmology (TTK), RWTH Aachen University, 52056 Aachen, Germany\label{aff50}
\and
Institut d'Estudis Espacials de Catalunya (IEEC),  Edifici RDIT, Campus UPC, 08860 Castelldefels, Barcelona, Spain\label{aff51}
\and
INAF-Osservatorio Astronomico di Roma, Via Frascati 33, 00078 Monteporzio Catone, Italy\label{aff52}
\and
INFN section of Naples, Via Cinthia 6, 80126, Napoli, Italy\label{aff53}
\and
Institute for Astronomy, University of Hawaii, 2680 Woodlawn Drive, Honolulu, HI 96822, USA\label{aff54}
\and
Dipartimento di Fisica e Astronomia "Augusto Righi" - Alma Mater Studiorum Universit\`a di Bologna, Viale Berti Pichat 6/2, 40127 Bologna, Italy\label{aff55}
\and
Institute for Astronomy, University of Edinburgh, Royal Observatory, Blackford Hill, Edinburgh EH9 3HJ, UK\label{aff56}
\and
Jodrell Bank Centre for Astrophysics, Department of Physics and Astronomy, University of Manchester, Oxford Road, Manchester M13 9PL, UK\label{aff57}
\and
European Space Agency/ESRIN, Largo Galileo Galilei 1, 00044 Frascati, Roma, Italy\label{aff58}
\and
Universit\'e Claude Bernard Lyon 1, CNRS/IN2P3, IP2I Lyon, UMR 5822, Villeurbanne, F-69100, France\label{aff59}
\and
Institut de Ci\`{e}ncies del Cosmos (ICCUB), Universitat de Barcelona (IEEC-UB), Mart\'{i} i Franqu\`{e}s 1, 08028 Barcelona, Spain\label{aff60}
\and
Instituci\'o Catalana de Recerca i Estudis Avan\c{c}ats (ICREA), Passeig de Llu\'{\i}s Companys 23, 08010 Barcelona, Spain\label{aff61}
\and
UCB Lyon 1, CNRS/IN2P3, IUF, IP2I Lyon, 4 rue Enrico Fermi, 69622 Villeurbanne, France\label{aff62}
\and
Mullard Space Science Laboratory, University College London, Holmbury St Mary, Dorking, Surrey RH5 6NT, UK\label{aff63}
\and
Departamento de F\'isica, Faculdade de Ci\^encias, Universidade de Lisboa, Edif\'icio C8, Campo Grande, PT1749-016 Lisboa, Portugal\label{aff64}
\and
Instituto de Astrof\'isica e Ci\^encias do Espa\c{c}o, Faculdade de Ci\^encias, Universidade de Lisboa, Campo Grande, 1749-016 Lisboa, Portugal\label{aff65}
\and
INFN-Padova, Via Marzolo 8, 35131 Padova, Italy\label{aff66}
\and
Aix-Marseille Universit\'e, CNRS/IN2P3, CPPM, Marseille, France\label{aff67}
\and
INFN-Bologna, Via Irnerio 46, 40126 Bologna, Italy\label{aff68}
\and
School of Physics, HH Wills Physics Laboratory, University of Bristol, Tyndall Avenue, Bristol, BS8 1TL, UK\label{aff69}
\and
FRACTAL S.L.N.E., calle Tulip\'an 2, Portal 13 1A, 28231, Las Rozas de Madrid, Spain\label{aff70}
\and
Dipartimento di Fisica "Aldo Pontremoli", Universit\`a degli Studi di Milano, Via Celoria 16, 20133 Milano, Italy\label{aff71}
\and
INFN-Sezione di Milano, Via Celoria 16, 20133 Milano, Italy\label{aff72}
\and
NRC Herzberg, 5071 West Saanich Rd, Victoria, BC V9E 2E7, Canada\label{aff73}
\and
Institute of Theoretical Astrophysics, University of Oslo, P.O. Box 1029 Blindern, 0315 Oslo, Norway\label{aff74}
\and
Jet Propulsion Laboratory, California Institute of Technology, 4800 Oak Grove Drive, Pasadena, CA, 91109, USA\label{aff75}
\and
Department of Physics, Lancaster University, Lancaster, LA1 4YB, UK\label{aff76}
\and
Felix Hormuth Engineering, Goethestr. 17, 69181 Leimen, Germany\label{aff77}
\and
Technical University of Denmark, Elektrovej 327, 2800 Kgs. Lyngby, Denmark\label{aff78}
\and
Cosmic Dawn Center (DAWN), Denmark\label{aff79}
\and
Institut d'Astrophysique de Paris, UMR 7095, CNRS, and Sorbonne Universit\'e, 98 bis boulevard Arago, 75014 Paris, France\label{aff80}
\and
Max-Planck-Institut f\"ur Astronomie, K\"onigstuhl 17, 69117 Heidelberg, Germany\label{aff81}
\and
NASA Goddard Space Flight Center, Greenbelt, MD 20771, USA\label{aff82}
\and
Department of Physics and Astronomy, University College London, Gower Street, London WC1E 6BT, UK\label{aff83}
\and
Department of Physics and Helsinki Institute of Physics, Gustaf H\"allstr\"omin katu 2, 00014 University of Helsinki, Finland\label{aff84}
\and
Universit\'e de Gen\`eve, D\'epartement de Physique Th\'eorique and Centre for Astroparticle Physics, 24 quai Ernest-Ansermet, CH-1211 Gen\`eve 4, Switzerland\label{aff85}
\and
Department of Physics, P.O. Box 64, 00014 University of Helsinki, Finland\label{aff86}
\and
Helsinki Institute of Physics, Gustaf H{\"a}llstr{\"o}min katu 2, University of Helsinki, Helsinki, Finland\label{aff87}
\and
Centre de Calcul de l'IN2P3/CNRS, 21 avenue Pierre de Coubertin 69627 Villeurbanne Cedex, France\label{aff88}
\and
Laboratoire d'etude de l'Univers et des phenomenes eXtremes, Observatoire de Paris, Universit\'e PSL, Sorbonne Universit\'e, CNRS, 92190 Meudon, France\label{aff89}
\and
Aix-Marseille Universit\'e, CNRS, CNES, LAM, Marseille, France\label{aff90}
\and
SKA Observatory, Jodrell Bank, Lower Withington, Macclesfield, Cheshire SK11 9FT, UK\label{aff91}
\and
Universit\"at Bonn, Argelander-Institut f\"ur Astronomie, Auf dem H\"ugel 71, 53121 Bonn, Germany\label{aff92}
\and
INFN-Sezione di Roma, Piazzale Aldo Moro, 2 - c/o Dipartimento di Fisica, Edificio G. Marconi, 00185 Roma, Italy\label{aff93}
\and
Department of Physics, Institute for Computational Cosmology, Durham University, South Road, Durham, DH1 3LE, UK\label{aff94}
\and
Infrared Processing and Analysis Center, California Institute of Technology, Pasadena, CA 91125, USA\label{aff95}
\and
Universit\'e C\^{o}te d'Azur, Observatoire de la C\^{o}te d'Azur, CNRS, Laboratoire Lagrange, Bd de l'Observatoire, CS 34229, 06304 Nice cedex 4, France\label{aff96}
\and
Universit\'e Paris Cit\'e, CNRS, Astroparticule et Cosmologie, 75013 Paris, France\label{aff97}
\and
CNRS-UCB International Research Laboratory, Centre Pierre Bin\'etruy, IRL2007, CPB-IN2P3, Berkeley, USA\label{aff98}
\and
University of Applied Sciences and Arts of Northwestern Switzerland, School of Engineering, 5210 Windisch, Switzerland\label{aff99}
\and
Institut d'Astrophysique de Paris, 98bis Boulevard Arago, 75014, Paris, France\label{aff100}
\and
Institute of Physics, Laboratory of Astrophysics, Ecole Polytechnique F\'ed\'erale de Lausanne (EPFL), Observatoire de Sauverny, 1290 Versoix, Switzerland\label{aff101}
\and
Aurora Technology for European Space Agency (ESA), Camino bajo del Castillo, s/n, Urbanizacion Villafranca del Castillo, Villanueva de la Ca\~nada, 28692 Madrid, Spain\label{aff102}
\and
Institut de F\'{i}sica d'Altes Energies (IFAE), The Barcelona Institute of Science and Technology, Campus UAB, 08193 Bellaterra (Barcelona), Spain\label{aff103}
\and
School of Mathematics, Statistics and Physics, Newcastle University, Herschel Building, Newcastle-upon-Tyne, NE1 7RU, UK\label{aff104}
\and
DARK, Niels Bohr Institute, University of Copenhagen, Jagtvej 155, 2200 Copenhagen, Denmark\label{aff105}
\and
Waterloo Centre for Astrophysics, University of Waterloo, Waterloo, Ontario N2L 3G1, Canada\label{aff106}
\and
Department of Physics and Astronomy, University of Waterloo, Waterloo, Ontario N2L 3G1, Canada\label{aff107}
\and
Perimeter Institute for Theoretical Physics, Waterloo, Ontario N2L 2Y5, Canada\label{aff108}
\and
Institute of Space Science, Str. Atomistilor, nr. 409 M\u{a}gurele, Ilfov, 077125, Romania\label{aff109}
\and
Consejo Superior de Investigaciones Cientificas, Calle Serrano 117, 28006 Madrid, Spain\label{aff110}
\and
Universidad de La Laguna, Departamento de Astrof\'{\i}sica, 38206 La Laguna, Tenerife, Spain\label{aff111}
\and
Caltech/IPAC, 1200 E. California Blvd., Pasadena, CA 91125, USA\label{aff112}
\and
Institut f\"ur Theoretische Physik, University of Heidelberg, Philosophenweg 16, 69120 Heidelberg, Germany\label{aff113}
\and
Institut de Recherche en Astrophysique et Plan\'etologie (IRAP), Universit\'e de Toulouse, CNRS, UPS, CNES, 14 Av. Edouard Belin, 31400 Toulouse, France\label{aff114}
\and
Universit\'e St Joseph; Faculty of Sciences, Beirut, Lebanon\label{aff115}
\and
Departamento de F\'isica, FCFM, Universidad de Chile, Blanco Encalada 2008, Santiago, Chile\label{aff116}
\and
Universit\"at Innsbruck, Institut f\"ur Astro- und Teilchenphysik, Technikerstr. 25/8, 6020 Innsbruck, Austria\label{aff117}
\and
Satlantis, University Science Park, Sede Bld 48940, Leioa-Bilbao, Spain\label{aff118}
\and
Centre for Electronic Imaging, Open University, Walton Hall, Milton Keynes, MK7~6AA, UK\label{aff119}
\and
Instituto de Astrof\'isica e Ci\^encias do Espa\c{c}o, Faculdade de Ci\^encias, Universidade de Lisboa, Tapada da Ajuda, 1349-018 Lisboa, Portugal\label{aff120}
\and
Cosmic Dawn Center (DAWN)\label{aff121}
\and
Niels Bohr Institute, University of Copenhagen, Jagtvej 128, 2200 Copenhagen, Denmark\label{aff122}
\and
Universidad Polit\'ecnica de Cartagena, Departamento de Electr\'onica y Tecnolog\'ia de Computadoras,  Plaza del Hospital 1, 30202 Cartagena, Spain\label{aff123}
\and
Dipartimento di Fisica e Scienze della Terra, Universit\`a degli Studi di Ferrara, Via Giuseppe Saragat 1, 44122 Ferrara, Italy\label{aff124}
\and
Istituto Nazionale di Fisica Nucleare, Sezione di Ferrara, Via Giuseppe Saragat 1, 44122 Ferrara, Italy\label{aff125}
\and
INAF, Istituto di Radioastronomia, Via Piero Gobetti 101, 40129 Bologna, Italy\label{aff126}
\and
Department of Physics, Oxford University, Keble Road, Oxford OX1 3RH, UK\label{aff127}
\and
Universit\'e PSL, Observatoire de Paris, Sorbonne Universit\'e, CNRS, LERMA, 75014, Paris, France\label{aff128}
\and
Universit\'e Paris-Cit\'e, 5 Rue Thomas Mann, 75013, Paris, France\label{aff129}
\and
INAF-Osservatorio Astronomico di Brera, Via Brera 28, 20122 Milano, Italy, and INFN-Sezione di Genova, Via Dodecaneso 33, 16146, Genova, Italy\label{aff130}
\and
ICL, Junia, Universit\'e Catholique de Lille, LITL, 59000 Lille, France\label{aff131}
\and
ICSC - Centro Nazionale di Ricerca in High Performance Computing, Big Data e Quantum Computing, Via Magnanelli 2, Bologna, Italy\label{aff132}
\and
Instituto de F\'isica Te\'orica UAM-CSIC, Campus de Cantoblanco, 28049 Madrid, Spain\label{aff133}
\and
CERCA/ISO, Department of Physics, Case Western Reserve University, 10900 Euclid Avenue, Cleveland, OH 44106, USA\label{aff134}
\and
Technical University of Munich, TUM School of Natural Sciences, Physics Department, James-Franck-Str.~1, 85748 Garching, Germany\label{aff135}
\and
Max-Planck-Institut f\"ur Astrophysik, Karl-Schwarzschild-Str.~1, 85748 Garching, Germany\label{aff136}
\and
Laboratoire Univers et Th\'eorie, Observatoire de Paris, Universit\'e PSL, Universit\'e Paris Cit\'e, CNRS, 92190 Meudon, France\label{aff137}
\and
Departamento de F{\'\i}sica Fundamental. Universidad de Salamanca. Plaza de la Merced s/n. 37008 Salamanca, Spain\label{aff138}
\and
Universit\'e de Strasbourg, CNRS, Observatoire astronomique de Strasbourg, UMR 7550, 67000 Strasbourg, France\label{aff139}
\and
Center for Data-Driven Discovery, Kavli IPMU (WPI), UTIAS, The University of Tokyo, Kashiwa, Chiba 277-8583, Japan\label{aff140}
\and
Ludwig-Maximilians-University, Schellingstrasse 4, 80799 Munich, Germany\label{aff141}
\and
Max-Planck-Institut f\"ur Physik, Boltzmannstr. 8, 85748 Garching, Germany\label{aff142}
\and
Dipartimento di Fisica - Sezione di Astronomia, Universit\`a di Trieste, Via Tiepolo 11, 34131 Trieste, Italy\label{aff143}
\and
NASA Ames Research Center, Moffett Field, CA 94035, USA\label{aff144}
\and
Bay Area Environmental Research Institute, Moffett Field, California 94035, USA\label{aff145}
\and
California Institute of Technology, 1200 E California Blvd, Pasadena, CA 91125, USA\label{aff146}
\and
Department of Physics \& Astronomy, University of California Irvine, Irvine CA 92697, USA\label{aff147}
\and
Department of Mathematics and Physics E. De Giorgi, University of Salento, Via per Arnesano, CP-I93, 73100, Lecce, Italy\label{aff148}
\and
INFN, Sezione di Lecce, Via per Arnesano, CP-193, 73100, Lecce, Italy\label{aff149}
\and
INAF-Sezione di Lecce, c/o Dipartimento Matematica e Fisica, Via per Arnesano, 73100, Lecce, Italy\label{aff150}
\and
Departamento F\'isica Aplicada, Universidad Polit\'ecnica de Cartagena, Campus Muralla del Mar, 30202 Cartagena, Murcia, Spain\label{aff151}
\and
Institute of Cosmology and Gravitation, University of Portsmouth, Portsmouth PO1 3FX, UK\label{aff152}
\and
Department of Computer Science, Aalto University, PO Box 15400, Espoo, FI-00 076, Finland\label{aff153}
\and
Instituto de Astrof\'\i sica de Canarias, c/ Via Lactea s/n, La Laguna 38200, Spain. Departamento de Astrof\'\i sica de la Universidad de La Laguna, Avda. Francisco Sanchez, La Laguna, 38200, Spain\label{aff154}
\and
Ruhr University Bochum, Faculty of Physics and Astronomy, Astronomical Institute (AIRUB), German Centre for Cosmological Lensing (GCCL), 44780 Bochum, Germany\label{aff155}
\and
Department of Physics and Astronomy, Vesilinnantie 5, 20014 University of Turku, Finland\label{aff156}
\and
Serco for European Space Agency (ESA), Camino bajo del Castillo, s/n, Urbanizacion Villafranca del Castillo, Villanueva de la Ca\~nada, 28692 Madrid, Spain\label{aff157}
\and
ARC Centre of Excellence for Dark Matter Particle Physics, Melbourne, Australia\label{aff158}
\and
Centre for Astrophysics \& Supercomputing, Swinburne University of Technology,  Hawthorn, Victoria 3122, Australia\label{aff159}
\and
Department of Physics and Astronomy, University of the Western Cape, Bellville, Cape Town, 7535, South Africa\label{aff160}
\and
DAMTP, Centre for Mathematical Sciences, Wilberforce Road, Cambridge CB3 0WA, UK\label{aff161}
\and
Kavli Institute for Cosmology Cambridge, Madingley Road, Cambridge, CB3 0HA, UK\label{aff162}
\and
Department of Astrophysics, University of Zurich, Winterthurerstrasse 190, 8057 Zurich, Switzerland\label{aff163}
\and
Department of Physics, Centre for Extragalactic Astronomy, Durham University, South Road, Durham, DH1 3LE, UK\label{aff164}
\and
IRFU, CEA, Universit\'e Paris-Saclay 91191 Gif-sur-Yvette Cedex, France\label{aff165}
\and
Oskar Klein Centre for Cosmoparticle Physics, Department of Physics, Stockholm University, Stockholm, SE-106 91, Sweden\label{aff166}
\and
Astrophysics Group, Blackett Laboratory, Imperial College London, London SW7 2AZ, UK\label{aff167}
\and
Univ. Grenoble Alpes, CNRS, Grenoble INP, LPSC-IN2P3, 53, Avenue des Martyrs, 38000, Grenoble, France\label{aff168}
\and
INAF-Osservatorio Astrofisico di Arcetri, Largo E. Fermi 5, 50125, Firenze, Italy\label{aff169}
\and
Dipartimento di Fisica, Sapienza Universit\`a di Roma, Piazzale Aldo Moro 2, 00185 Roma, Italy\label{aff170}
\and
Centro de Astrof\'{\i}sica da Universidade do Porto, Rua das Estrelas, 4150-762 Porto, Portugal\label{aff171}
\and
HE Space for European Space Agency (ESA), Camino bajo del Castillo, s/n, Urbanizacion Villafranca del Castillo, Villanueva de la Ca\~nada, 28692 Madrid, Spain\label{aff172}
\and
Department of Astrophysical Sciences, Peyton Hall, Princeton University, Princeton, NJ 08544, USA\label{aff173}
\and
Theoretical astrophysics, Department of Physics and Astronomy, Uppsala University, Box 515, 751 20 Uppsala, Sweden\label{aff174}
\and
Minnesota Institute for Astrophysics, University of Minnesota, 116 Church St SE, Minneapolis, MN 55455, USA\label{aff175}
\and
Mathematical Institute, University of Leiden, Einsteinweg 55, 2333 CA Leiden, The Netherlands\label{aff176}
\and
School of Physics \& Astronomy, University of Southampton, Highfield Campus, Southampton SO17 1BJ, UK\label{aff177}
\and
Institute of Astronomy, University of Cambridge, Madingley Road, Cambridge CB3 0HA, UK\label{aff178}
\and
Department of Physics and Astronomy, University of California, Davis, CA 95616, USA\label{aff179}
\and
Space physics and astronomy research unit, University of Oulu, Pentti Kaiteran katu 1, FI-90014 Oulu, Finland\label{aff180}
\and
Center for Computational Astrophysics, Flatiron Institute, 162 5th Avenue, 10010, New York, NY, USA\label{aff181}}    

\date{\today}

\abstract{The star-forming main sequence (SFMS) is a tight relation observed between stellar masses and star formation rates (SFR) in a population of galaxies. This relation is observed at different redshifts, in various morphological, and environmental domains, and is key to understanding the underlying relations between a galaxy budget of cold gas and its stellar content. \Euclid Quick Data Release 1 (Q1) gives us the opportunity to investigate this fundamental relation in galaxy formation and evolution. We complement the \Euclid release with public IRAC observations of the \Euclid Deep Fields, improving the quality of recovered photometric redshifts, stellar masses, and SFRs, as is shown both with simulations and a comparison with available spectroscopic redshifts. From Q1 data alone, we recover more than $\sim 30\,\mathrm{k}$ galaxies with $\Mstarwun > 11$, giving a precise constraint of the SFMS at the high-mass end. We investigated the SFMS, in a redshift interval between $0.2$ and $3.0$, comparing our results with the existing literature and fitting them with a parameterisation taking into account the presence of a bending of the relation at the high-mass end, depending on the bending mass, $M_0$. We find good agreement with previous results in terms of $M_0$ values, and an increasing trend for the relation scatter at higher stellar masses. We also investigate the distribution of physical (e.g. dust absorption, $A_V$, and formation age) and morphological properties (e.g., Sérsic index and radius) in the SFR--stellar mass plane, and their relation with the SFMS. These results highlight the potential of \Euclid in studying the fundamental scaling relations that regulate galaxy formation and evolution in anticipation of the forthcoming Data Release 1.}


\titlerunning{\Euclid\/: the SFMS in the EDFs}
\authorrunning{Euclid Collaboration: Enia et al.}
   
\maketitle   
\section{\label{sc:Intro}Introduction}
The \gls{ms} is a relation between stellar masses ($M_{\ast}$) and \gls{sfr} that is observed for \gls{sfg}. It has been extensively studied in the last few decades \citep{2004MNRAS.351.1151B, 2007ApJ...670..156D, 2011A&A...533A.119E}, with investigations into its slope, normalisation, scatter, and evolution over time \citep[see][and references therein]{2014ApJS..214...15S, 2023MNRAS.519.1526P}. The \gls{ms} is observed across different redshifts and is already in place by $z \sim 6$ \citep[e.g.,][]{2024ApJ...977..133C, 2025ApJ...979..193C}. It hosts the majority of star formation at each epoch \citep{2011ApJ...739L..40R}, suggesting that galaxies spend most of their lifetimes on the \gls{ms}, undergoing secular evolution. The tightness of the relation, with a typical scatter between $0.2\,\mathrm{dex}$ and $0.4\,\mathrm{dex}$, implies its universality as the main mode of galaxy growth \citep{2015A&A...575A..74S, 2019MNRAS.485.4817D, 2019MNRAS.484..915M}.

This relation emerges from the interplay between the stellar content of galaxies and their cold gas reservoirs (\citealp[i.e., the so-called Kennicutt--Schmidt relation from][]{1959ApJ...129..243S} \citealp[and][]{1998ARA&A..36..189K}, \citealp[and the (resolved or integrated) molecular gas main-sequence, see e.g.][]{2019ApJ...884L..33L, 2020MNRAS.496.4606M, 2021MNRAS.501.4777E, 2023MNRAS.518.4767B}), and it has been shown to hold at sub-kiloparsec scales too \citep[e.g.,][]{2013ApJ...779..135W, 2017ApJ...851L..24H, 2017ApJ...851...18L, 2017MNRAS.469.2806A, 2018MNRAS.474.2039E, 2020MNRAS.493.4107E, 2022MNRAS.510.3622B}.

At masses of $\Mstarwun > 10$ at $0<z<1$ and at the high-mass end $\Mstarwun > 11$ at $z\sim2$, the relation appears to exhibit a deviation from the linear trend, the so-called bending of the \gls{ms} \citep{2014ApJ...795..104W, 2015A&A...575A..74S, 2016ApJ...817..118T, 2019MNRAS.483.3213P, 2022ApJ...936..165L, 2022A&A...661L...7D, 2024A&A...691A.248L, 2024A&A...681A.110W}. The bending traces changes in cold-gas accretion \citep{2005MNRAS.363....2K, 2006MNRAS.368....2D} and availability for star formation processes, and could be a consequence of the reduced availability of cold gas in halos entering the hot accretion mode phase \citep{2022A&A...661L...7D}, or feedback from active galactic nuclei \citep{2012ARA&A..50..455F}, or both \citep{2017MNRAS.465...32B}. Additionally, the reactivation of star formation in the disks of galaxies that are approaching quiescence or have already been quenched may also contribute to the bending of the \gls{ms} \citep{2019MNRAS.489.1265M}. This turnover mass can be linked with the host halo mass quenching threshold \citep{2007ApJ...671..153Y, 2019MNRAS.488.3143B, 2020ApJ...899...58L, 2023MNRAS.519.1526P}, defining the transition between an environment favourable to star formation to a regime where these processes are suppressed.

The \cite{Q1cite} is the first release of \Euclid survey data, corresponding to a single reference observing sequence \citep[ROS, see][]{Scaramella-EP1} of the \gls{EDF}. This is a homogeneous view of a large area of the extragalactic sky ($\sim 63\,\mathrm{deg}^2$) from optical to \gls{NIR}, complemented with observations of the same fields at $3.6\,\micron$ and $4.5\,\micron$ with the Infrared Array Camera \citep[IRAC,][]{2004ApJS..154...10F} on {\it Spitzer} \citep{2004ApJS..154....1W}. These fields have the potential to become the best-studied extragalactic fields of the coming decades. In this work, we illustrate the results obtained with the data and products of Q1 for the \gls{ms}, investigating its evolution up to $z = 3$, and the distribution of physical and morphological parameters along the \gls{ms}, as well as validating these results with the existing literature, a first demonstration of the potential of \Euclid to investigate scaling relations and the baryon cycle.

This paper is structured as follows. In Sect.\,\ref{sc:Data}, we describe the data released for Q1. In Sect.\,\ref{sc:Methodology}, we describe the methods used to recover the photometric redshifts and \gls{pp}. In the same section, we validate the results, reporting the performance of our methods on simulations and the available sub-sample of spectroscopic redshifts and H$\alpha$-estimated SFRs. In Sect.\,\ref{sc:Results}, we report the results for the \gls{ms}. In Sect.\,\ref{sc:Conclusions}, we present our conclusions and perspectives for the upcoming Data Release 1.

Throughout this paper we adopt a flat Lambda cold dark matter ($\Lambda$CDM) cosmology with $H_0 = 70\,\kmsMpc$, $\Omega_{\rm m} = 0.3$, and $\Omega_{\Lambda} = 0.7$, and assume a \citet{2003PASP..115..763C} initial mass function (IMF). All magnitudes are given in the AB photometric system \citep{1983ApJ...266..713O}.

\section{\label{sc:Data}Data}
A detailed description of the Q1 data release is presented in \citet{Q1-TP001}, \citet{Q1-TP002}, \citet{Q1-TP003}, and \citet{Q1-TP004}. A summary of the scientific objectives of the mission can be found in \cite{EuclidSkyOverview}. In short, for Q1 \Euclid observed $\sim 63\,\mathrm{deg}^2$ of the extragalactic sky, divided into EDF-North (EDF-N), EDF-Fornax (EDF-F), and EDF-South (EDF-S), in four photometric bands, one in the visible \citep[$\IE$,][]{EuclidSkyVIS}, and three in the \gls{NIR} \citep[$\YE$, $\JE$, and $\HE$,][]{EuclidSkyNISP}. These observations are complemented by ground-based observations carried out with multiple instruments to cover the wavelength range between $0.3\,\micron$ and $1.8\,\micron$ by the Ultraviolet Near-Infrared Optical Northern Survey \citep[UNIONS,][]{2025arXiv250313783G} and the Dark Energy Survey \citep[DES,][]{2015AJ....150..150F, 2016MNRAS.460.1270D}.

In order to obtain robust results and improve the quality of the recovered photometric redshifts and \gls{pp} (see Sect.\,\ref{sc:Validation}), we also added to the \Euclid photometry two available IRAC bands, at $3.6\,\micron$ and $4.5\,\micron$, covering all the \gls{EDF} \citep{Moneti-EP17, EP-McPartland}. More details on how IRAC photometry is measured can be found in \citet{Q1-SP011}.

In Table\;\ref{tab:filters} we report the filters used in this work, with the observed $10\sigma$ depths for an extended source in an aperture that is twice the full width at half maximum \citep[FWHM, i.e., the one with the lowest resolution between the optical and \Euclid bands, see][for further details]{Q1-TP004}. For this work, we started from the available \Euclid catalogues and applied a series of selections to make our analysis more robust, removing compact or low-quality sources. These selections are:
\begin{itemize}
    \item \texttt{SPURIOUS$\_$FLAG} $= 0$;
    \item \texttt{DET\_QUALITY\_FLAG} $< 4$;
    \item \texttt{MUMAX\_MINUS\_MAG} $> -2.6$.
\end{itemize}
For further details on the meaning of the flags, see \citet{Q1-TP005}. This selection skims the sample from stars and compact objects such as quasi-stellar objects (QSOs).
\begin{table}[]
    \centering
    \caption{Filters used in this work, with associated observed depths.}\label{tab:filters}
    \begin{tabular}{lccccc}
        \hline
        \hline
        \noalign{\vskip 1pt}
        Band &  $\lambda_{\rm eff}$ [\micron] & EDF-N & EDF-F & EDF-S \\
        \hline
        \noalign{\vskip 1pt}
        $u_\sfont{CFHT/MegaCam}$ & $0.372$ & $23.39$ & $     $ & $     $ \\
        $g_\sfont{HSC}$          & $0.480$ & $24.87$ & $     $ & $     $ \\
        $r_\sfont{CFHT/MegaCam}$ & $0.640$ & $24.01$ & $     $ & $     $ \\
        $i_\sfont{PAN-STARRS}$   & $0.755$ & $23.07$ & $     $ & $     $ \\
        $z_\sfont{HSC}$          & $0.891$ & $23.35$ & $     $ & $     $ \\
        $g_\sfont{Decam}$        & $0.473$ & $     $ & $24.65$ & $24.72$ \\
        $r_\sfont{Decam}$        & $0.642$ & $     $ & $24.33$ & $24.37$ \\
        $i_\sfont{Decam}$        & $0.784$ & $     $ & $23.76$ & $23.78$ \\
        $z_\sfont{Decam}$        & $0.926$ & $     $ & $23.06$ & $23.12$ \\
        VIS/\IE                  & $0.715$ & $24.75$ & $24.70$ & $24.74$ \\
        NISP/\YE                 & $1.085$ & $23.16$ & $23.10$ & $23.15$ \\ 
        NISP/\JE                 & $1.375$ & $23.31$ & $23.24$ & $23.30$ \\
        NISP/\HE                 & $1.773$ & $23.24$ & $23.19$ & $23.24$ \\
        {\rm IRAC1}              & $3.550$ & $24.05$ & $24.05$ & $23.15$ \\
        {\rm IRAC2}              & $4.493$ & $23.95$ & $23.95$ & $23.05$ \\
        \hline
    \end{tabular}
    \tablefoot{Reported magnitudes are the $10\sigma$ observed median depths of the observing tiles for an extended source in a $2 \times \mathrm{FWHM}$ diameter aperture. For IRAC values see \citet{Moneti-EP17} and \citet{EP-McPartland}.}
\end{table}

We further cleaned our sample from these two classes of objects using the classification probability of the Q1 data products, imposing the following criteria:
\begin{itemize}
    \item \texttt{PROB$\_$QSO} $< 0.86$;
    \item \texttt{PROB$\_$STAR} $< 0.10$.
\end{itemize}
See Sect.\,4 of \citet{Q1-TP005} for further details, and also \citet{Q1-SP027} about the classification thresholds.

Finally, we benefited from the results of the morphological analysis for Q1 \citep{Q1-SP047, Q1-SP040}, applying another set of cuts related to the morphological parameters and the size of the source. We kept sources with:
\begin{itemize}
    \item $q > 0.05$;
    \item $0.01a < R_{\mathrm e} < 2a$.
\end{itemize}
$q$ is the Sérsic axis ratio, $R_{\mathrm e}$ the Sérsic radius, and $a$ the isophote semi-major axis, in units of VIS pixels. For further information see Sect.\,4 and the left panel of Fig.\,3 in \citet{Q1-SP040}. These further remove diffraction spikes, cosmic rays, or stars that survived the cuts described above.

The last cut that we applied was in magnitude, in order to work with a mass-complete sample, limiting our analysis to sources with observed $\HE < 24$, corresponding to an average measured signal-to-noise ratio of five, measured from the $2\times\,\mathrm{FWHM}$ aperture photometry. Our final sample is composed of \num{8090074} sources.

\section{\label{sc:Methodology}Physical properties and validation}
We refer the reader to \citet{Q1-TP005} for a description of how the Q1 data were processed to infer photometric redshifts and \gls{pp}. In this section, we briefly summarise the procedure.
\begin{figure*}
    \centering
    \includegraphics[width=\hsize]{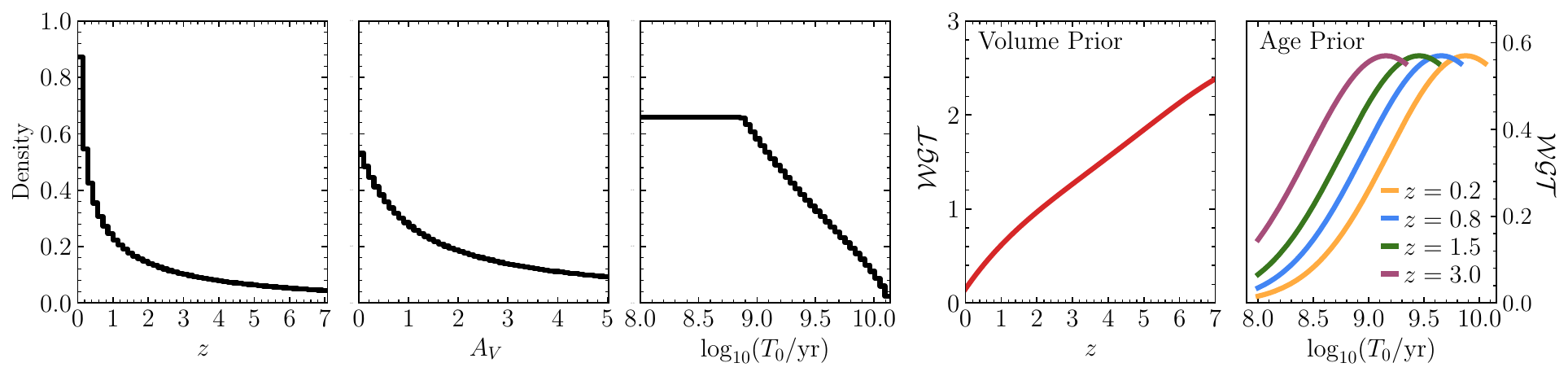}
    \caption{\emph{Left panels}: Distribution of redshifts, $z$, attenuations, $A_v$, and ages, $T_0$ for the reference sample used in this work. \emph{Right panels}: Two priors adopted in this work -- volume and age -- as multiplicative weights on the reference galaxies, depending on their redshifts or ages.}
    \label{fig:refsample}
\end{figure*}

Due to the large number of sources detected (of the order of tens of millions) in the \gls{EDF}, \gls{ml} methods have been developed to speed up the computational process while achieving a comparable performance of template fitting methods \citep[see e.g.,][]{Desprez-EP10, EP-Enia}. Data products produced for Q1 have been obtained with a \gls{nn} algorithm, \nnpz, which finds a $k$ number of \gls{nn} (30 in the \Euclid pipeline, 80 for this work) in the template space (i.e., magnitude and colour) for each target galaxy from a reference sample, and infers the photometric redshifts and \gls{pp} from those.

The reference sample differs from the one used to produce the Q1 data products, the reason for which is explained below. It was built from a grid of templates \citep[][2016 version\footnote{\url{http://www.bruzual.org/bc03/Updated_version_2016/}}]{2003MNRAS.344.1000B}, using the MILES stellar library, which adopts the \citet{2001MNRAS.322..231K} IMF, which we therefore converted to \citet{2003PASP..115..763C} taking into account a difference of about 0.03 dex for \gls{pp}. The models were built with exponentially delayed star formation histories:
\begin{equation}
    \mathrm{SFR} (t) \propto (t-T_0)\,\mathrm{e}^{-(t-T_0)/\tau},
\end{equation}
drawn from a \citet{Halton} grid\footnote{A method to generate a quasi-random grid, which is evenly distributed across the parameter space, minimising the presence of gaps and clusters of points.} in a six-dimensional space with the following free parameters: 
\begin{itemize}
    \item redshifts: $0 < z < 7$;
    \item ages: $8 < \log_{10}(T_0/\mathrm{yr}) < 10.138$;
    \item e-folding timescale: $8 < \log_{10}(\tau/\mathrm{yr}) < 10.5$;
    \item ionisation parameter: $-4 < \log_{10} (U) < -2$;
    \item metallicities: $0.1 < Z/Z_\odot < 2$.
\end{itemize}
For dust attenuation, we generated models with both \citet{Calzetti2000} and SMC \citep{1984A&A...132..389P} laws, with $V$-band attenuation, $A_V$, between $0$ and $5$. Stellar masses and \gls{sfr} were inferred from the amplitude of the observed \gls{sed}, as a scaling parameter recovered by \nnpz. Metallicities and ionisation parameters are distributed uniformly within the given ranges; the same for $\tau$, but on a logarithmic scale. The redshifts are distributed with a linear scaling in $(1+z)$ steps, while ages and $A_V$ scale logarithmically from the lowest value to the highest. For these last three parameters, their distribution in the reference sample is shown in black in the first three panels of Fig.\,\ref{fig:refsample}. We then generated the noise-free observed-frame photometry associated with each model. In the end, the reference sample consists of \num{1490150} objects.

\nnpz\ then finds the $80$ \gls{nn} for every target galaxy, based on the observed magnitudes and colours. Each of these \gls{nn} will have its own weight -- measured from the $\chi^2$ distance between the reference and the target -- and scaling parameter from the \gls{sed} amplitude. By combining them, we measure the median (or the mode) of the distribution of \gls{nn}, which ultimately are the inferred photo-$z$s, \gls{pp}, and absolute magnitudes of the target galaxies.

Unlike what has been done in the \Euclid pipeline, we added two more photometric points to the reference sample, accounting for the IRAC1 and IRAC2 channels (see Sect.\,\ref{sc:Data}). Moreover, having access to the \nnpz\ results -- that is, the set of \gls{nn} for each target galaxy -- we can impose whatever physically motivated condition (i.e. a prior) directly on the \gls{nn}, either by measuring the output, $z_{\rm phot}$ or \gls{pp}, only on the set of \gls{nn} that satisfy the condition, or by differently re-weighting the \gls{nn} in the reference sample in order to penalise unphysical solutions.

As is reported in Sect.\,6.1.1 of \citet{Q1-TP005}, the Q1 pipeline results (obtained without the application of any condition to the \gls{nn}) contain an artificially high number of low-$z$ galaxies with extremely young ages -- $\log_{10}(T_0/\mathrm{yr})$ starts from $7$ in the pipeline -- observed at the peak of their star-forming activity, and thus at the limit of \gls{sSFR} inherent to parametric models \citep[$\log_{10} \mathrm{sSFR/\mathrm{yr}^{-1}} \simeq -7.8$, see Fig.\,8 of][]{2017A&A...608A..41C}. The resulting redshift distribution is ultimately skewed towards non-physically high number counts at low $z$, creating an artefact straight line at the \gls{sSFR} saturation limit \citep[see Fig.\,14 in][]{Q1-TP005}. In principle, this issue could be addressed in the Q1 pipeline data product by imposing an age prior on the \gls{nn}; for example, ignoring those with ages $< 0.1\,\mathrm{Gyr}$. This would significantly reduce the impact of artefacts, but at the cost of reducing the Q1 sample by about $40\%$, excluding from the sample all sources without even a single NN that satisfies the prior. Losing these sources would introduce systematic biases into our analysis. To avoid this, we took some precautions that deviate from what has been done in the pipeline, in order to reduce the impact of those artefact sources without losing a significant fraction of the sample: the boundaries for ages and $A_\mathrm{V}$ mentioned above used to generate the reference sample for this work are different from those reported in \citet{Q1-TP005}, with $0 < A_\mathrm{V} < 3$ and $7 < \log_{10}(T_0/\mathrm{yr}) < 10.138$. We also increased the $k$ number of \gls{nn} to $80$. Finally, we imposed both a volume and an age prior on the \gls{nn} in the reference sample.
\begin{figure*}
    \centering
    \includegraphics[width=\hsize]{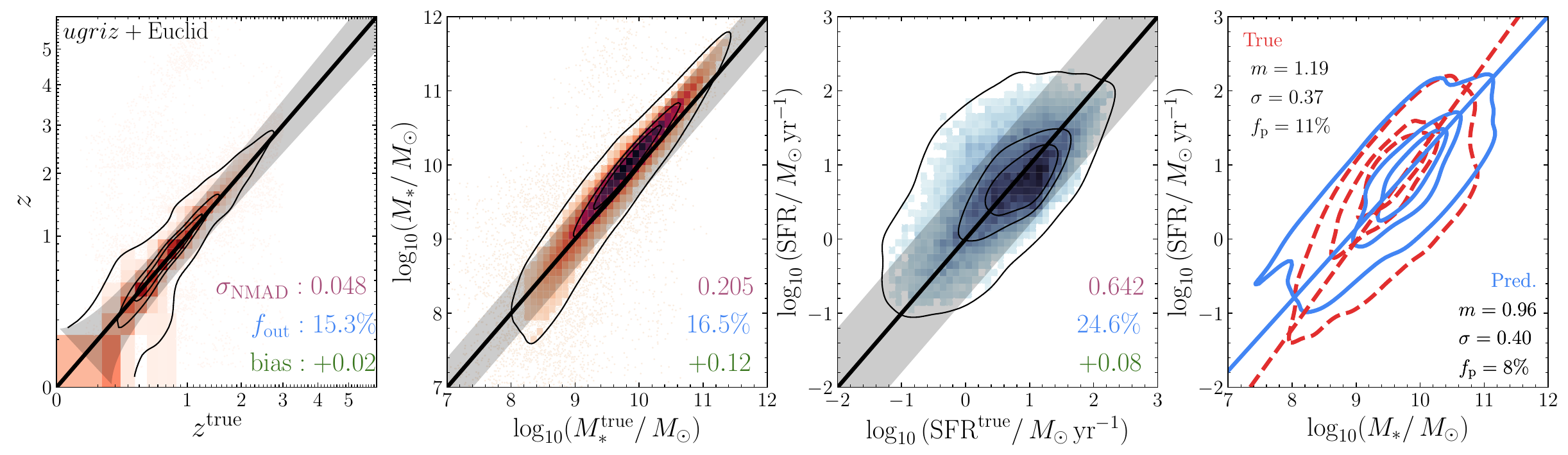}
    \includegraphics[width=\hsize]{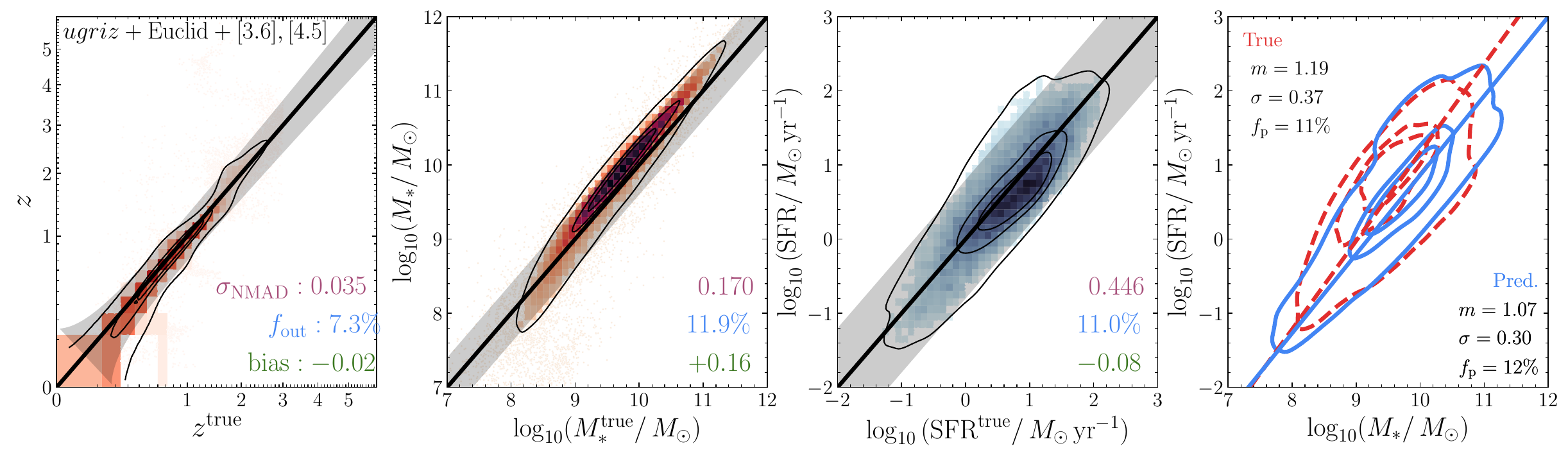}
    \caption{Results for the \nnpz\ run on the $70\,\mathrm{k}$ simulated FS2 galaxies, without (\emph{top panels}) and with the first two WISE filters (\emph{bottom panels}). The black line is the 1:1 relation; the shaded area is the region beyond which a prediction is an outlier. In every panel, the contours are the area containing $86\%$, $39\%$ (corresponding to the $2\sigma$ and $1\sigma$ levels for a 2D histogram), and $20\%$ of the sample. For \gls{ms} (right panel), the true distribution is reported in red (dashed) and the predicted one in blue (solid). The lines are the \gls{odr} best fit to the distribution with passive galaxies removed. The reported metrics are NMAD (purple), the outlier fraction, $f_{\rm out}$ (blue), and the bias (green) for the photometric redshifts and physical parameters, and the slope, $m$, scatter, $\sigma$, and fraction of passive galaxies, $f_{\rm p}$, for the \gls{ms}, all defined in Sect.\,\ref{sc:metrics}.}
    \label{fig:FS2_results}
\end{figure*}

With the volume prior, we increased the weights of \gls{nn} at higher redshifts compared to the ones at lower redshifts, where the volume of the Universe sampled by the survey is smaller and fewer galaxies are expected to be observed. This prior was implemented as a multiplicative weight assigned to each object in the reference sample, and only depends on the redshift as
\begin{equation}
\mathcal{WGT}(z) \propto \frac{{\rm d}V_{\rm c}(z)}{{\rm d}z},
\label{eq:volprior}
\end{equation}
where ${\rm d}V_{\rm c}(z)$ is the comoving volume shell in the interval $[z,\, z+\mathrm{d}z]$. This prior is shown in red in the centre-right panel of Fig.\,\ref{fig:refsample}.

The age prior is once again a multiplicative weight, $\mathcal{WGT}(T_0)$, to apply to \gls{nn}. This takes into account the fact that younger galaxies could be observed at higher redshifts, while this possibility should be reduced at low redshifts. In building the prior, we looked at the distribution of ages in different redshift bins in the $2\,\mathrm{deg}^2$
of the Cosmic Evolution Survey \citep[COSMOS,][]{2022ApJS..258...11W}, finding that these can be modelled as normal distributions with the peak age decreasing while moving at a higher redshift. This weight is then constructed as a truncated normal distribution centred on two thirds the age of the Universe at any given $z$, with width $0.7$ in $\log_{10}(T_0/\mathrm{yr})$. A weight equal to zero is assigned for those sources with an age greater than the age of the Universe at the given $z$. The shape of the age prior is reported in the rightmost panel of Fig.\,\ref{fig:refsample} for four indicative redshift values.

This procedure makes us sensitive to the bulk of the population of \gls{sfg}, while reducing our ability to properly identify and describe outliers (e.g., the starburst galaxies, as the presence of a star-forming burst is not directly accounted for in the reference sample). Given the main scopes of this work, this is acceptable, since it has been found that the exponentially delayed $\tau$ model is an accurate description of the star formation histories of the main population of \gls{ms} galaxies, at least at $z < 2$ \citep[see e.g.,][]{2014ApJS..214...15S, 2017A&A...608A..41C}. Recently, studies focussing on non-parametric models of star formation histories have found how these models could better recover the complex events that arise during the evolution of a galaxy \citep{2019ApJ...879..116I, 2019ApJ...876....3L, 2020A&A...641A.119B}. In the future, for Data Release 1, models will better account for starburst galaxies, with the possibility of exploring other regions of the parameter space, even accounting for complex star formation histories \citep[see for example][]{Q1-SP044}. 
\begin{figure*}
    \centering
    \includegraphics[width=\hsize]{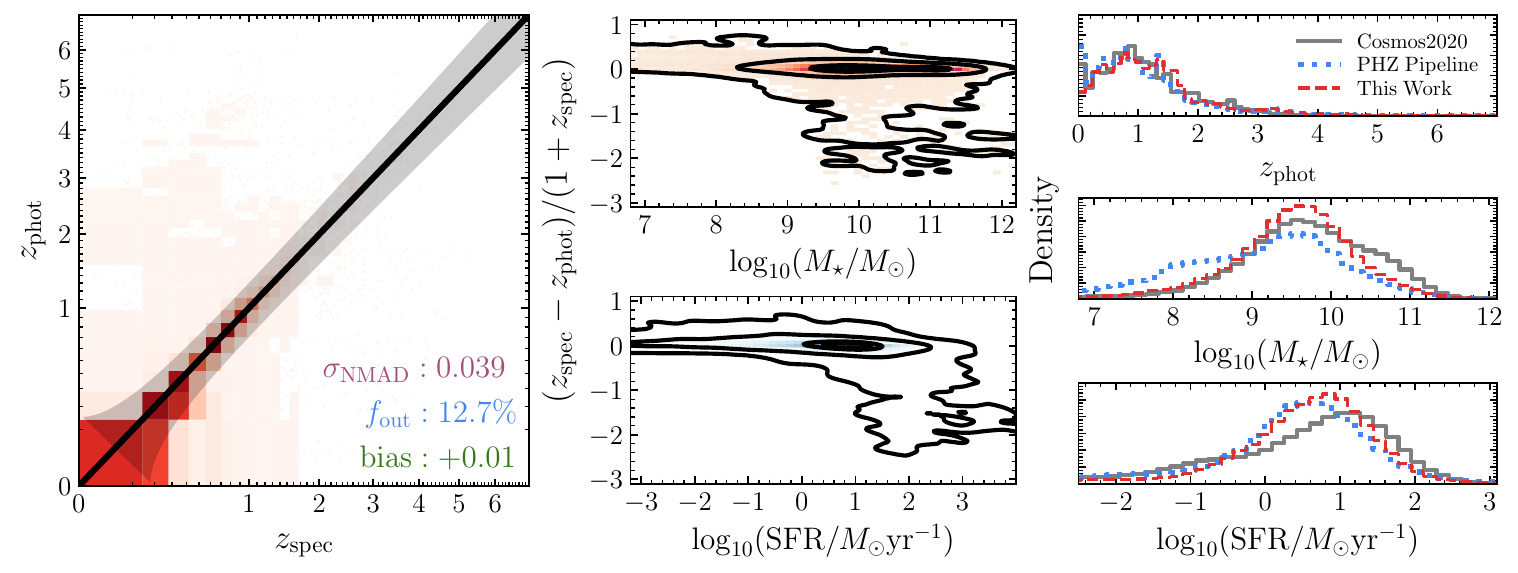}
    \caption{\emph{Left}: Comparison between the measured photometric redshifts and a compilation of all the available and reliable spectroscopic redshifts, colour-coded by the density of objects in each bin. The black line is the 1:1 relation, while the shaded grey area is the region beyond which a prediction is considered to be an outlier. The reported metrics are NMAD (purple), the outlier fraction, $f_{\rm out}$ (blue), and the bias (green), all defined in Sect.\,\ref{sc:metrics}. \emph{Middle}: Normalised redshift difference as a function of stellar masses (top) and \gls{sfr} (bottom). The black contours are at the same levels as in Fig.\,\ref{fig:FS2_results}. \emph{Right}: Normalised distributions of photometric redshifts (top), stellar masses (centre), and \gls{sfr} (bottom) for our full sample (dashed red lines), compared with the results coming from the Q1 data products (PHZ, in dotted blue lines), and COSMOS2020 (in solid grey lines), at the same magnitude cuts applied in this work (see Sect.\,\ref{sc:Data}).}
    \label{fig:z_comparison}
\end{figure*}
\begin{figure*}
    \centering
    \includegraphics[width=\hsize]{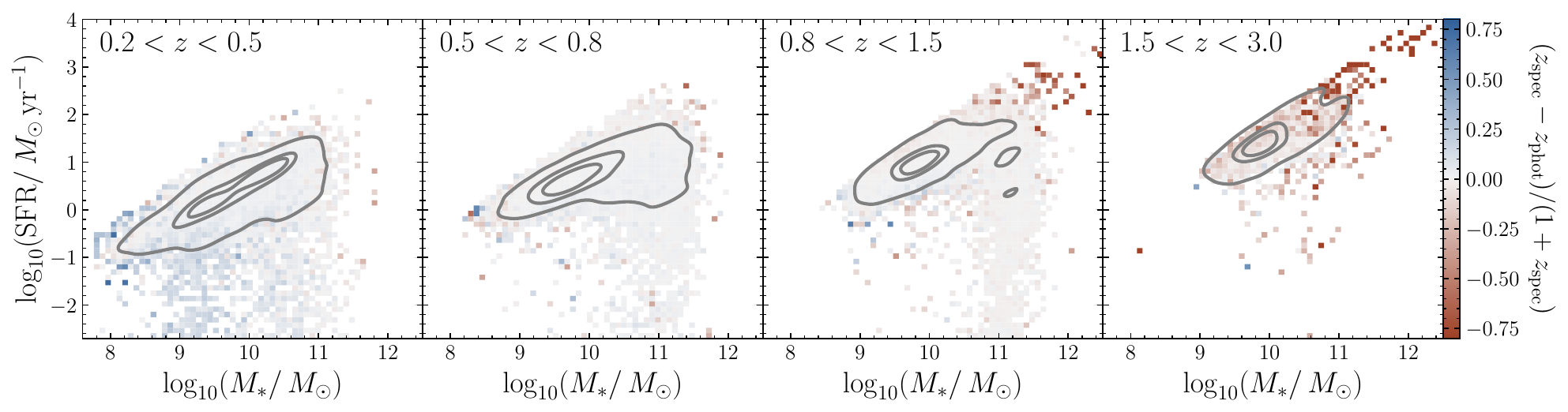}
    \caption{$M_\ast$--SFR plane at four different redshift intervals, only for the sources with a reliable $z_{\rm spec}$ described in Sect.\,\ref{sc:Validation}, colour-coded by the median normalised redshift difference in each bin. To give an idea of the density of objects in each redshift interval, we superimposed the contours in grey, with the same levels as in Fig.\,\ref{fig:FS2_results}.}
    \label{fig:MS_vs_deltaz}
\end{figure*}
\begin{figure*}
    \sidecaption
    \centering
    \includegraphics[angle=0,width=12.9cm]{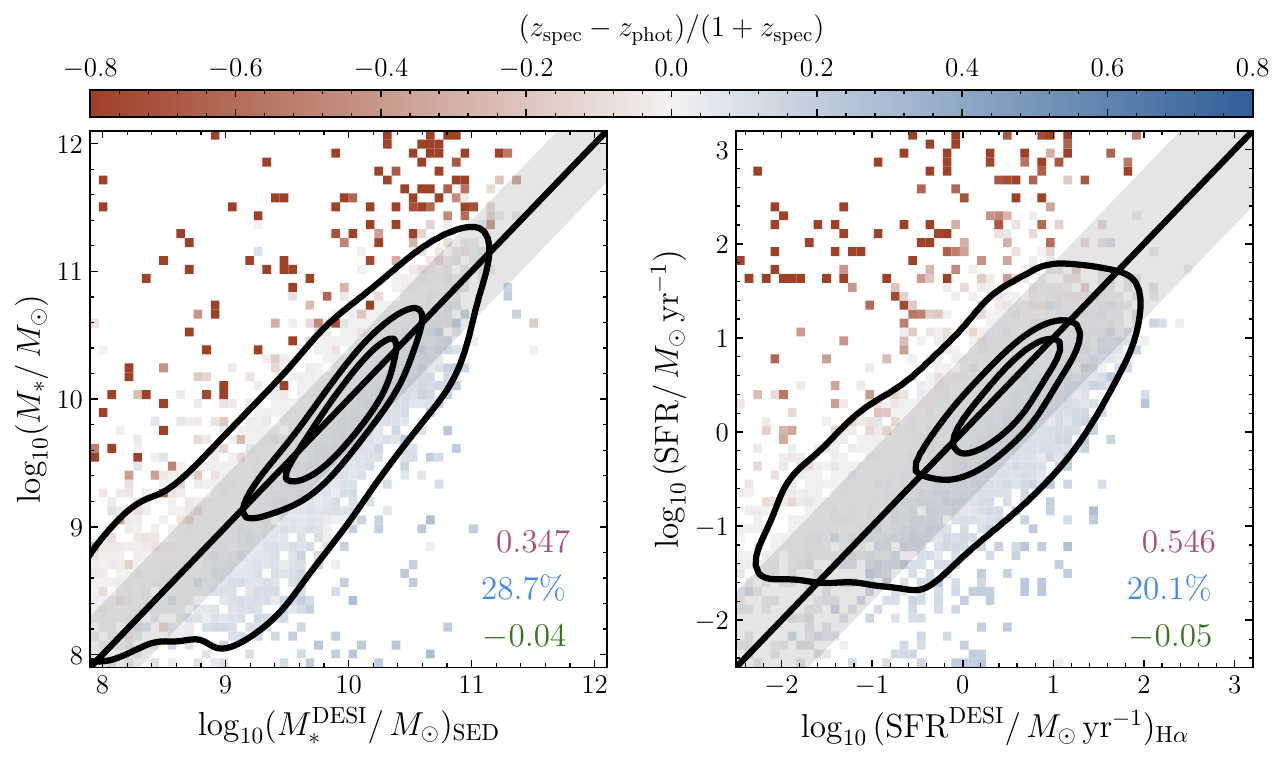}
    \caption{Comparison between the stellar masses and \gls{sfr} measured in this work and a sample of objects from DESI with stellar masses from SED fitting (\emph{left}) and \gls{sfr} from H$\alpha$ (\emph{right}), colour-coded as a function of the median normalised redshift difference in each bin. The black line is the 1:1 relation, while the shaded area is the region beyond which a prediction is an outlier. The contours are the same as in Fig.\,\ref{fig:FS2_results}. The reported metrics are defined in Sect.\,\ref{sc:metrics}.}
    \label{fig:DESI_val}
\end{figure*}

We validated our results both on the available state-of-the-art simulations adapted to reproduce as closely as possible \Euclid observations \citep[i.e., the Flagship2 simulation, FS2, see][]{EuclidSkyFlagship}, and on a compilation of the available spectroscopic redshifts and H$\alpha$ measured \gls{sfr} in the \gls{EDF}. Although the latter is fundamental to assess the quality of recovered parameters with what can be interpreted as the closest possible thing to a ‘ground truth’ value, the former is necessary to put a degree of confidence in all the other recovered quantities that have no such ‘ground truth’ to compare with.
\begin{figure}
    \centering
    \includegraphics[width=0.9\hsize]{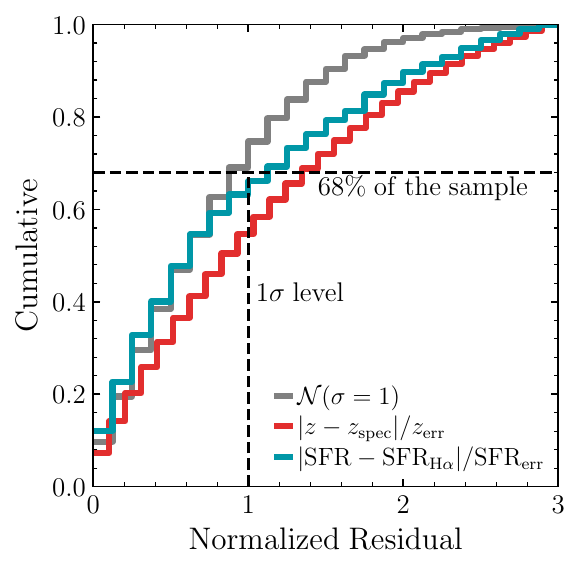}
    \caption{Cumulative distribution of the normalised residuals for photometric redshifts (red) and \gls{sfr} (blue). For comparison, we also show in grey the cumulative of a normal distribution with $\sigma = 1$. Dashed black lines highlight the position where $68\%$ of the distributions are, and the $1\sigma$ level for the residuals.}
    \label{fig:uncertainties}
\end{figure}

\subsection{Metrics for quality assessment}\label{sc:metrics}
The metrics used to quantify the quality of the results are defined differently when referring to redshifts or \gls{pp}. We refer the reader to \cite{EP-Enia} for a full discussion of thresholds and catastrophic outlier definitions; here, we simply report the definitions.

We first defined a set of true values, $z_{\rm test}$ and $y_{\rm test}$ (on a logarithmic scale for \gls{pp}), to confront with the predicted values, $z_{\rm pred}$ and $y_{\rm pred}$. We then defined the normalised median absolute deviation as
\begin{equation}
{\rm NMAD} = 1.48\,\times\,{\rm median}
\begin{cases}
 \,\dfrac{|z_{\rm pred} - z_{\rm test}|}{1+z_{\rm test}} - b, & \text{for redshifts,} \\
 \,|y_{\rm pred} - y_{\rm test}| - b, & \text{for \gls{pp},}
\end{cases}
\end{equation}
with $b$ being the model bias (see below).

Then, we defined the outlier fraction,
\begin{equation}
f_{\rm out}\,=
    \begin{cases}
    \,\dfrac{|z_{\rm pred} - z_{\rm test}|}{1+z_{\rm test}} > 0.15, & \text{for redshifts,} \\
    \,|y_{\rm pred} - y_{\rm test}| > t_{\rm out}, & \text{for \gls{pp},} 
    \end{cases}
\end{equation}
with $t_{\rm out} = 0.4$ for stellar masses and $t_{\rm out} = 0.8$ for \gls{sfr}.

Finally, we defined the bias,
\begin{equation}
b\,={\rm median}
    \begin{cases}
    \,\left( \dfrac{z_{\rm pred} - z_{\rm test}}{1+z_{\rm test}} \right), & \text{for redshifts,} \\
    \,(y_{\rm pred} - y_{\rm test}), & \text{for \gls{pp}.}
    \end{cases}
\end{equation}

\subsection{Validation on simulations}\label{sc:simulations}
We randomly selected about $70\,\mathrm{k}$ sources from a complete octant of the FS2 simulation \citep{EuclidSkyFlagship}. We checked that the selection does not skew the ground-truth values with respect to the full distribution. These mock sources are, by construction, distributed in the redshift range $0 < z < 3$, with $6.4 < \Mstarwun < 12$ and $-4 < \sfrwun < 3.1$.  We then followed the same procedure described in Sect.\,2 of \citet{EP-Enia} to produce a realistic catalogue of mock sources that reproduces as closely as possible what is observed in Q1. We perturbed the intrinsic fluxes with the noise level found in Q1 in the same set of filters (see Table\;\ref{tab:filters}), and cut at observed magnitudes $\HE < 24$. Instead of IRAC, in the FS2 simulation these wavelength ranges are covered by mock observations with WISE bands, W1 and W2 \citep{2010AJ....140.1868W}. Given the similar range covered, we ensured that the resulting performance is comparable if not exactly the same.

We then ran \nnpz\ with the same reference sample described in Sect.\,\ref{sc:Methodology}, one run with the WISE filters and one without, and produced the recovered results as the median of the 80 \gls{nn}. The results are shown in Fig.\,\ref{fig:FS2_results}, where we report the recovered versus true relation without (top panel) and with (bottom panel) the two WISE filters. These results follow closely what has already been described in \citet{EP-Enia}, at least at the order-of-magnitude level. It is immediately noticeable how the addition of the two filters at $3.6\,\micron$ and $4.5\,\micron$ improves parameter estimation, with photo-$z$ NMADs and the outlier fraction decreasing from $0.048$ to $0.035$ and from $15\%$ to $7\%$, respectively. The same holds for stellar masses (NMAD decreasing from $0.206$ to $0.170$ and the outlier fraction from $17\%$ to $12\%$) and especially for \gls{sfr} (NMAD decreasing from $0.643$ to $0.446$ and the outlier fraction from $25\%$ to $11\%$). As was expected, the recovery of \gls{ms} improves with respect to the case without the two filters in \gls{NIR}, with a much better recovery of the slope and normalisation of the \gls{ms} relation.

\subsection{Photometric redshifts and PPs validation}\label{sc:Validation}
Quality assessment for the redshifts was performed by looking at how they compare with respect to the observed spectroscopic ones. The \gls{EDF} cover regions of the sky where there is a plethora of coverage from other spectroscopic surveys. In total, we successfully matched \num{63504} galaxies with a reliable spectroscopic redshift; that is, with a redshift quality flag of 3 or 4 \citep[see description in Sect. 5.2 of][]{Q1-TP005}. These $z_{\rm spec}$ values are from: the Dark Energy Spectroscopic Instrument \citep[DESI,][]{2016arXiv161100036D, 2024AJ....168...58D}; the 16th Data Release of the Sloan Digital Sky Survey \citep[SDSS,][]{2020ApJS..249....3A}; the 2MASS Redshift Survey \citep[2MRS,][]{2012ApJS..199...26H}; the PRIsm MUlti-object Survey \citep[PRIMUS,][]{2011ApJ...741....8C}; the Australian Dark Energy Survey \citep[OzDES,][]{2015MNRAS.452.3047Y, 2017MNRAS.472..273C, 2020MNRAS.496...19L}; 3dHST \citep{2012ApJ...758L..17B}; the 2-degree Field Galaxy Redshift Survey \citep[2dFGRS,][]{2001MNRAS.328.1039C}; the 6-degree Field Galaxy Redshift Survey \citep[6dFGS,][]{2009MNRAS.399..683J}; the MOSFIRE Deep Evolution Field Survey \citep[MOSDEF,][]{2015ApJS..218...15K}; the VANDELS ESO public spectroscopic survey \citep{2018arXiv181105298P, 2023A&A...678A..25T}; the JWST Advanced Deep Extragalactic Survey DR3 \citep[JADES,][]{2025ApJS..277....4D}; the 2-degree Field Lensing Survey \citep[2dFLens,][]{2016MNRAS.462.4240B}; and the VIMOS VLT deep survey \citep[VVDS,][]{2005A&A...439..845L}.

The results are reported in Fig.\,\ref{fig:z_comparison}. In the left panel, we compare the photometric and spectroscopic redshifts for the subset of reliable spectroscopic redshifts in our sample. These are almost equally divided between EDF-F and EDF-N, with only a select number (\num{123}) of objects in EDF-S. The trend we find using \Euclid real data is similar to what is shown in the previous section, with a non-negligible improvement when adding the two IRAC bands. To put it in quantitive terms, compared to the same analysis performed without the two IRAC bands (not shown in Fig.\,\ref{fig:z_comparison}) the NMAD decreases from $0.06$ to $0.04$ and the fraction of outliers decreases from $26\%$ to $13\%$. In the central panels, we show the normalised difference between spectroscopic and photometric redshifts, as a function of the stellar masses and the \gls{sfr}. We do not observe any troubling systematic trend, with the exception of the (expected) behaviour in which galaxies mistakenly placed at higher redshift -- mostly those with $(z_{\rm spec} - z_{\rm phot})/(1+z_{\rm spec}) > -1$ -- are found with an incorrect higher \gls{sfr}, so some caution must be taken when dealing with those high-$z$ galaxies, or with $\sfrwun > 2.4$. In terms of observed bias, we find a negligible value of $+0.01$, well below the typical uncertainties associated with photo-$z$s (see below).

When looking at the full sample of objects -- not just spectroscopic ones -- there are no ground-truth values to compare with, but we can still investigate how our results agree with the full distribution of photometric redshifts, stellar masses, and \gls{sfr}, especially when compared with other surveys. This is done in the right panels in Fig.\,\ref{fig:z_comparison}, where we compare our results for the full sample (the dashed red lines) with what is observed in the LePhare run from the classic catalog of COSMOS 2020 (the solid grey lines), and with the inferred \gls{pp} from the Q1 data products (PHZ, the dotted blue lines), both at the same magnitude cut applied in this work (i.e. $\HE < 24$). The first two distributions in redshift are comparable, with the main differences observed in a slightly lower fraction of $0.6 < z < 1.2$ objects in our sample (and conversely, a few more $z > 3$ galaxies), while the Q1 data products exhibit a significantly greater number of $z < 0.2$ objects. As for the stellar masses, our results improve with respect to the almost flat at $8 < \Mstarwun < 9$ distributions of Q1 data products; however, we find fewer objects in the $10.2 < \Mstarwun < 11.2$ range and conversely more in the $9.2 < \Mstarwun < 10.2$ range with respect to COSMOS. The biggest differences are observed in terms of \gls{sfr}, with more galaxies in our sample in the $0 < \sfrwun < 1$ range with respect to COSMOS, and conversely slightly fewer galaxies per SFR bin in our sample at $1 < \sfrwun < 2$.

In Fig.\,\ref{fig:MS_vs_deltaz}, we report all this information in the $M_\ast$--SFR plane, where the \gls{ms} is observed. For the sub-sample of sources with a reliable spectroscopic redshift, we show the median value of the normalised redshift difference in each bin, with red colours highlighting sources mistakenly placed at a higher redshift, and blue colours the opposite. While the latter catastrophic outliers are a small issue only visible in the first redshift interval (with $0.2 < z < 0.5$), the former becomes more and more prominent at higher redshifts (i.e., $z>1.5$), introducing a non-negligible bias in the estimates of $M_\ast$ and SFR in the highest mass and SFR regimes, with these skewed towards higher values due to the wrong redshift attribution.

We compare the measured \gls{pp} with the ones in the value-added catalogue of \gls{pp} in DESI \citep{2024A&A...691A.308S}, although for a limited number of sources ($\sim 5\,\mathrm{k}$) in the interval $0 < z < 0.6$. In particular, we confront the \gls{sfr} with the ones obtained from both H$\alpha$ and H$\beta$ line measurement -- not exactly ‘ground truth’ values, but close -- while the stellar masses are compared to their SED fitting results, even though, given how it has been performed on fewer and shallower filters \citep[see Table 1 in][]{2024A&A...691A.308S}, we expect our stellar masses to be more reliable. The IMF is the same as the one adopted for this work \citep{2003PASP..115..763C}, and \gls{sfr} were obtained from H$\alpha$ following \citet{1998ApJ...498..541K}. The results are shown in Fig.\,\ref{fig:DESI_val}, where the result for each PP is colour-coded as a function of the median value of $(z_{\rm spec} - z_{\rm phot})/(1+z_{\rm spec})$ in each bin. Despite the limited sample, the performance is in line with what is expected from the simulations. It is immediately clear how the vast majority of catastrophic outliers (i.e. where \gls{pp} fall outside of the defined thresholds) are a consequence of sources with an incorrect photometric redshift estimate. Again, the observed bias in stellar masses and SFR is still below the typical uncertainties associated with each \gls{pp}.

Finally, we used these \gls{sfr} from H$\alpha$ and the reliable spectroscopic redshift sample to place some constraints on the estimated uncertainties in the photo-$z$s and \gls{sfr}. Uncertainties were measured from the $16$th and $84$th percentiles of the weighted distribution of \gls{nn} (see Sect.\,\ref{sc:Methodology}). To account for the possibility of an under- (or over-) estimation of those uncertainties, we looked at the cumulative distribution of $(z_{\rm phot} - z_{\rm spec})/z_{\rm err}$ -- and similarly $(\mathrm{SFR} - \mathrm{SFR}_{\rm H \alpha})/\mathrm{SFR}_{\rm err}$ -- where we expect $68\%$ of these to fall below $1$ if the uncertainties are well estimated. In contrast, an underestimate would lead to fewer sources within the $1\sigma$ limit, and the opposite would be true for an overestimate. These cumulative distributions (‘normalised residuals’, red for redshifts and blue for \gls{sfr}) are reported in Fig.\,\ref{fig:uncertainties}. We find that our uncertainties are slightly underestimated for redshifts ($1\sigma$ level reached for $55\%$ of the sample) and almost spot on for \gls{sfr}. We estimate the underestimation of the uncertainties of the photometric redshifts to be a factor of about $1.5$.

We take the mode of the distribution of uncertainties as the typical values for each parameter, which are $0.05$ for redshifts, $0.11$ for stellar masses, and $0.08$ for \gls{sfr} (on a logarithmic scale). The typical uncertainty on \gls{sfr} are used in the following section for the fit of the \gls{ms}.
 
\section{\label{sc:Results}Results}
We started from the sample described in Sect.\,\ref{sc:Data}, and limited our analysis to the redshift range between $0.2$ and $3.0$. This is motivated by the need to obtain a statistically reliable sample in terms of mass completeness and the quality of the recovered photometric redshifts and physical properties. In \citet{EP-Enia}, it is shown how the main source of biases in the analysis of \gls{ms} during cosmic time -- apart from the inherent dispersions in determining the correct \gls{pp} -- arises from the photo-$z$ estimation, whereby typically some low-$z$ (i.e. $z < 0.5$) objects are placed at high $z$ (up to around $10\%$ of catastrophic outliers) with increased stellar masses and \gls{sfr}. The net effect is a steepening of the \gls{ms} at lower redshifts, and the opposite at higher $z$ \citep[see Fig.\,11 of][]{EP-Enia}. This is also observed in the validation tests that we performed with simulations and reported in Sect.\,\ref{sc:Validation} (see Fig.\,\ref{fig:MS_vs_deltaz}).
\begin{figure}
    \centering
    \includegraphics[width=\hsize]{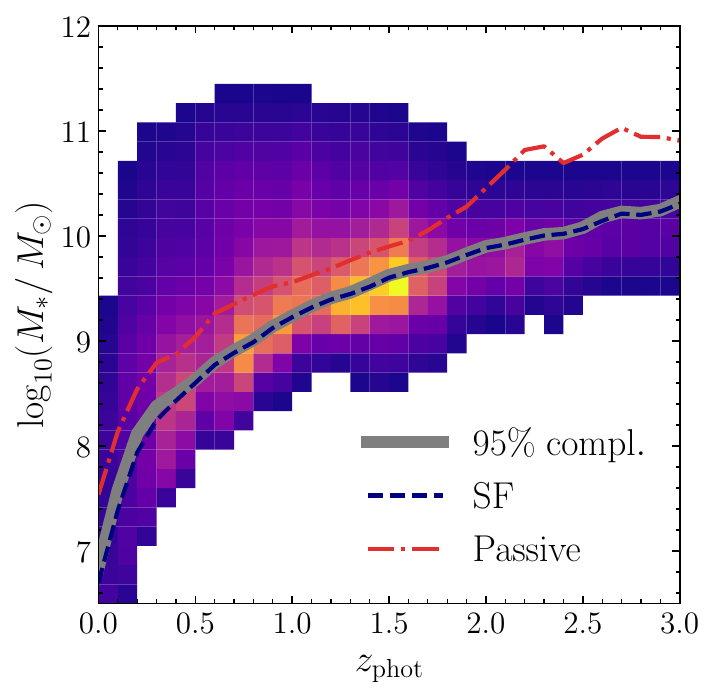}
    \caption{The stellar mass completeness for the sample used in this work \citep{2010A&A...523A..13P}, with a cut at $\HE < 24$, shown as the density of galaxies in the photometric redshifts vs. stellar masses plane. The solid grey line is the $95\%$ stellar mass completeness limit, red dashed-dotted line for passive galaxies, and blue  dashed line for \gls{sfg}.}
    \label{fig:masscompl}
\end{figure}

We performed our subsequent analysis in the following redshift bins: $[0.2$, $0.5], [0.5$, $0.8], [0.8$, $1.5], [1.5$, $3.0]$, with the exception of the morphological analysis, where we stopped at $z = 1.5$, since for higher redshifts the quality of the recovered morphological parameters is limited by the sizes of the sources reaching the resolution limit of the survey. Based on the results shown in Fig.\,\ref{fig:MS_vs_deltaz}, we can place a certain degree of confidence in the highest-mass end of the first two redshift intervals, while greater caution is required for the last two. In each redshift bin, we limited our analysis -- and the reported values -- to $\Mstarwun < 11.5$.
\begin{figure*}
    \centering
    \includegraphics[width=\hsize]{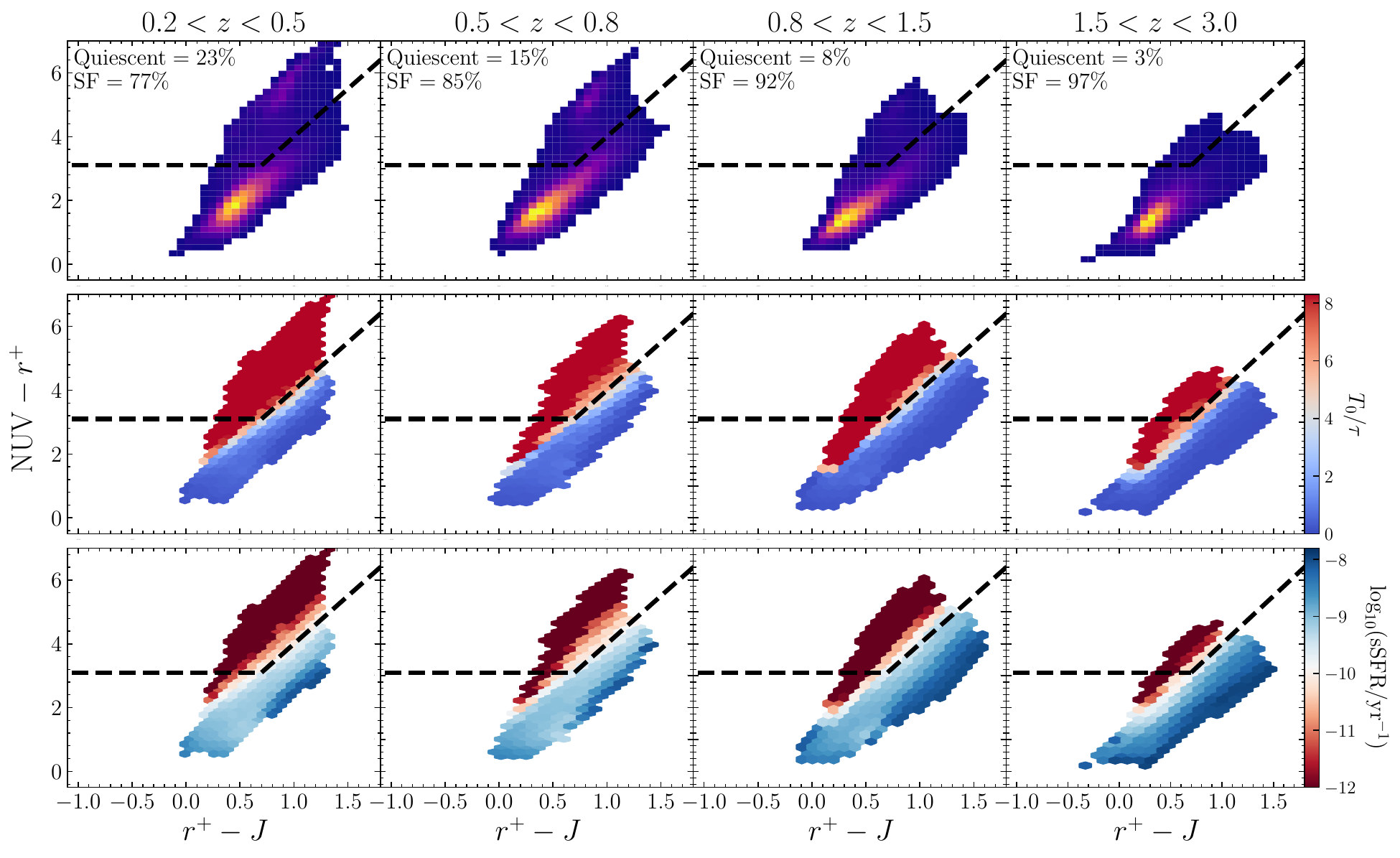}
    \caption{$\mathrm{NUV}-r^{+}$ versus $r^{+}-J$ rest-frame colours for the sources above the $95\%$ mass completeness limit, in four redshift bins, colour-coded with the density of objects (\emph{upper panels}), the ratio between $T_0$ and $\tau$ (\emph{middle panels}), and the \gls{sSFR} (\emph{bottom panels}). Dash-dotted lines divide the regions between star-forming and quiescent galaxies, following the criteria presented in \citet{2013A&A...556A..55I}.}
    \label{fig:NUVRJ_full}
\end{figure*}

\subsection{Mass completeness}
We estimated the mass completeness of the sample limited to $\HE < 24$ following the method described in \citet{2010A&A...523A..13P}. We selected galaxies close to the limiting magnitude of our sample; that is, those with $23 < \HE < 24$ \citep[with identical results of choosing the faintest $20\%$ as in][]{2010A&A...523A..13P}, and measured for each galaxy
\begin{equation}
\log_{10} (M_{\rm lim}/M_\odot) = \Mstarwun - 0.4(\HE - 24),
\end{equation}
representing the mass the galaxies would have at the limiting magnitude. We then measured the $95$th percentile of the distribution of $M_{\rm lim} (z)$ for each redshift bin. This is reported in Fig.\,\ref{fig:masscompl}. The classification into star-forming (dashed blue line) and passive galaxies (dash-dotted red line) was done with the selection criteria based on the $\mathrm{NUV}$--$r^+$--$J$ diagram (as is explained below).

The sample is around $95\%$ complete for $M_{\ast}$, which increases from $\Mstarwun \sim 7$ (at $z \sim 0.1$) to $\sim 9$ (at $z \sim 1$), and increases from $\Mstarwun \sim 9.5$ to $\Mstarwun \sim 10$, while going to higher redshifts, from $z = 1.5$ to $2.5$. For passive galaxies, the limit is $0.3$ -- $0.4\,\mathrm{dex}$ higher up to $z \sim 1.8$, and about $0.7\,\mathrm{dex}$ higher at $z > 2$. Up to $z = 3$, there are \num{4654430} galaxies over the $95\%$ stellar mass completeness limit.

\subsection{Star-forming and passive galaxy classification}
Colour-based classifications of galaxies use the principle of separating red and blue galaxies and distinguish between dusty and intrinsically red ones \citep[e.g. the $U-V$ versus $V-J$ colour diagram][]{2005ApJ...624L..81L, 2007ApJ...655...51W, 2010ApJ...713..738W}. In this work, we use the $\mathrm{NUV}-r^+$, $r^+-J$ colours \citep{2010ApJ...709..644I}, whereby star-forming and passive galaxies are discriminated based on their absolute $\mathrm{NUV}$, $r^{+}$, and $J$ magnitudes, with quiescent galaxies satisfying the following relations:
\begin{eqnarray}
\mathrm{NUV}-r^+&>&3(r^+-J)+1; \nonumber \\
\mathrm{NUV}-r^+&>&3.1.
\end{eqnarray}
This combination of criteria works in a similar fashion to the $UVJ$ diagram, but it has been widely used in recent literature as it is more sensitive to recent star formation via the $\mathrm{NUV} - r^+$ colour, which separates passive (redder) and star-forming (bluer) galaxies \citep{2013A&A...556A..55I, 2013A&A...558A..67A}. Truly passive galaxies are then distinguished from dusty, star-forming objects via the $r^+ - J$ colour, ensuring a proper separation between the two different populations.
\begin{figure*}
    \centering
    \includegraphics[width=\hsize]{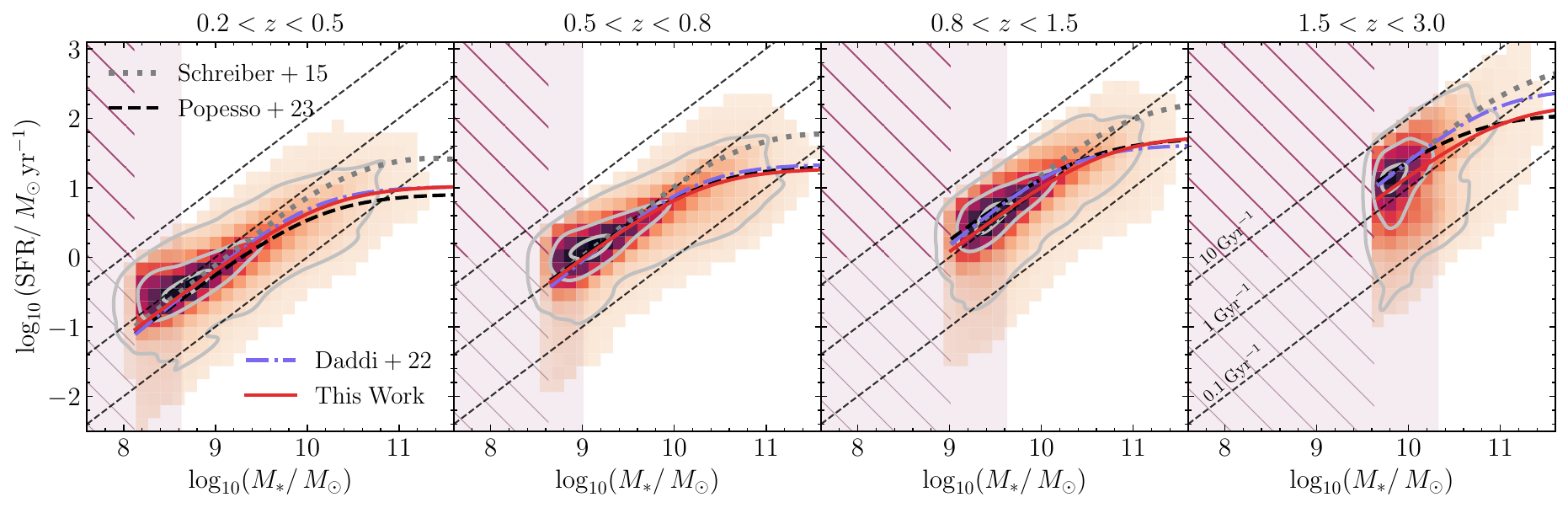}
    \caption{$\mathrm{SFR}$--$M_\ast$ relation for \gls{sfg}, selected with the $\mathrm{NUV}-r^{+}$ versus $r^{+}-J$ condition, coloured by the density of objects in each redshift bin. Within the coloured bins is $99.9\%$ of the sample of \gls{sfg}, while the grey contours are the lines enclosing $95\%$, $50\%$, and $10\%$ of it. The red line is the fitted \gls{ms} presented in this work, starting from the completeness limit mass at the start of the $z$-bin. The other \gls{ms}s are from \citet[][dashed black line]{2023MNRAS.519.1526P}, \citet[][dotted grey line]{2015A&A...575A..74S}, and \citet[][dashed-dotted blue line]{2022A&A...661L...7D}, all evaluated at the median redshift of each redshift bin (or to the closest reported value). The shaded purple regions highlight where the mass is below the completeness limit at the start (hatched) and at the end of the redshift bin. Dashed grey lines highlight the region of constant sSFR.}
    \label{fig:MS_full}
\end{figure*}

The colours $\mathrm{NUV}-r^{+}$ versus $r^{+}-J$ of our sample are shown in Fig.\,\ref{fig:NUVRJ_full}, colour-coded as a function of the number density of objects with certain colours (top panels) and the median $T_0/\tau$ (middle panels) and \gls{sSFR} in each bin (bottom panels), in the four different redshift bins. Only objects whose mass is higher than the $M_{\lim} (z)$ value at the $z_{\rm min}$ of the redshift bin are shown, to account for sample incompleteness (see Fig.\,\ref{fig:masscompl}). To be consistent with the selection criteria reported above, the adopted absolute magnitude, $J$, is not the one estimated in the rest frame \Euclid $\JE$ filter but in the $J_\sfont{UltraVISTA}$ filter \citep{Schirmer-EP18}; similarly, we adopted the $r^+$ band as in \citet{2010ApJ...709..644I}. 

We recover the well-known increase in the fraction of passive galaxies while going to later cosmic times, highlighting the assembly of the population of quiescent galaxies observed in the Universe. When checking the colour diagrams against the measured values of \gls{sSFR}, we notice the presence of a small fraction of objects ($\sim 3\%$) outside the boundaries for quiescent galaxies in the $\mathrm{NUV}$--$r^+$--$J$ diagram, but with a median value of $\log_{10}(\mathrm{sSFR}/\mathrm{yr}^{-1}) < -11$ (and $T_0/\tau > 8$); that is, in a region of the $M_\ast$--SFR plane where we would expect only quiescent galaxies. This is a small, negligible number of interlopers, which is reassuring with regard to the goodness of the classification based on the rest-frame colours.

\subsection{The SFR--$M_\ast$ relation in the EDFs}\label{sc:MS}
The $M_\ast$--SFR plane for the sample of \gls{sfg} selected with the $\mathrm{NUV}$--$r^{+}$--$J$ colours is shown in Fig.\,\ref{fig:MS_full}, in four redshift bins, colour-coded in terms of the logarithmic density of objects per bin. The \gls{ms} is observed up to the highest redshift bin ($z < 3$). In the figure, we report three previously published \gls{ms} relations for comparison: Eq.\;(15) of \citet[][black dashed line]{2023MNRAS.519.1526P}, that is, a comprehensive compilation of 27 literature \gls{ms} relations fitted to the same functional form, in the mass range $8.5 < \Mstarwun < 11.5$; \citet[][dotted grey line]{2015A&A...575A..74S}, obtained from $\sim 10\,\mathrm{k}$ galaxies with the deepest {\it Herschel} observations of the GOODS and CANDELS-Herschel programme, with $9.5 < \Mstarwun < 11.5$; and \citet[][dash-dotted blue line]{2022A&A...661L...7D}, obtained from a stacking analysis of $\sim 400\,\mathrm{k}$ colour-selected \gls{sfg} in COSMOS \citep{2021A&A...647A.123D}, here too in the mass range $8.5 < \Mstarwun < 11.5)$.

All these \gls{ms} forms include the presence of a bending of the relation at the high-mass end. Having enough statistics in terms of the number of objects per bin, we can significantly constrain the deviation from the linear form at high masses. For example, in each redshift interval \citet{2021A&A...647A.123D} and \citet{2022A&A...661L...7D} found no more than $838$ \gls{sfg} at $\Mstarwun > 11$, and in the same mass range the stacking analysis in \citet{2015A&A...575A..74S} has $< 15$ galaxies per bin. Analogously, most of the studies in the compilation of \citet{2023MNRAS.519.1526P} stop at $\Mstarwun = 11$, and those that extend further never reach more than $10^3$ galaxies per bin in comparable redshift ranges. Due to the large area observed in Q1, the minimum number of galaxies we have at $\Mstarwun > 11$ is $875$, in the $[0.2$, $0.5]$ redshift bin. This number increases to $7640$ (for $0.5 < z < 0.8$), $18951$ (for $0.8 < z < 1.5$), and $8165$ (for $1.5 < z < 3.0$), one or two orders of magnitude higher than the former reported statistics. These numbers are more than enough to subdivide the $\Mstarwun > 11$ region into two different bins at $z > 0.8$, to better constrain the part of the \gls{ms} where the SFR appears to saturate. The same reasoning, scaled by an order of magnitude, applies if we consider $\Mstarwun > 10.5$.

Our fit to the observed data is the red line in Fig.\,\ref{fig:MS_full}. Overall, the fits in \citet{2023MNRAS.519.1526P} and \citet{2022A&A...661L...7D} closely resemble our results, while \citet{2015A&A...575A..74S} find systematically higher \gls{sfr} at the highest mass end. The functional form that we fit these data to, first proposed by \citet{2015ApJ...801...80L}, is the same as in Eq.\;(15) of \citet{2023MNRAS.519.1526P} or Eq.\;(1) of \citet{2022A&A...661L...7D},
\begin{equation}\label{eq:MSform}
    \mathrm{SFR} = \frac{\mathrm{SFR}_{\rm max}}{1+(M_0/M_\ast)^{\gamma}};
\end{equation}
that is, a parameterisation where the SFR is linked to the stellar mass through three parameters -- the bending mass after which the relation deviates from the linear behaviour ($M_0$), the maximum SFR for $M_{\ast} \gg M_0$ ($\mathrm{SFR}_\mathrm{max}$), and the slope of the linear relation when $M_{\ast} \ll M_0$ ($\gamma$). These parameters have been shown to be directly linked to fundamental properties in models of gas accretion \citep[e.g. the bending mass $M_0$ is directly linked to the ratio between $M_{\rm shock}$ and $M_{\rm stream}$, see][]{2022A&A...661L...7D, 2022ApJ...926L..21D}. For each fit, we kept $\gamma$ fixed at $1.0$, as it has been shown to be a representative value around which almost every fit converges at different redshifts \citep[see][]{2015ApJ...801...80L, 2023MNRAS.519.1526P, 2022A&A...661L...7D}. This also makes it possible to make a direct comparison with these works in terms of bending mass and the maximum SFR.
\begin{figure}
    \centering
    \includegraphics[width=\hsize]{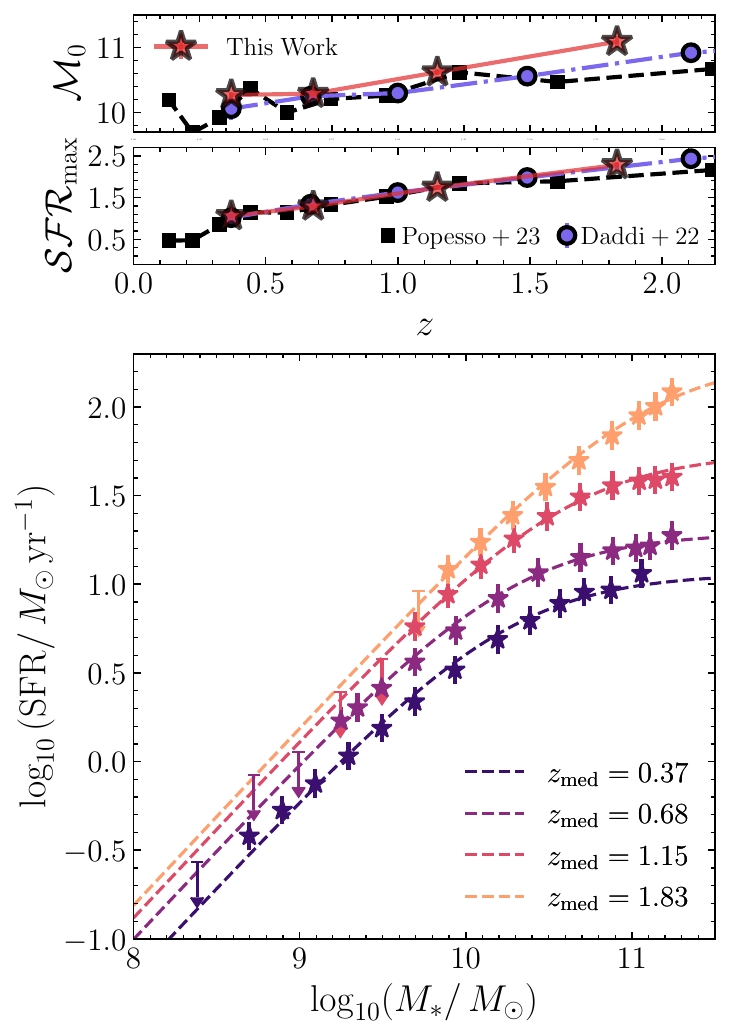}
    \caption{\emph{Lower panel}: Our fit of the \gls{ms}, in the four redshift bins (stars with error bars and arrows) and best-fit model (dashed lines). The reported redshift is the median $z$ of the bin. \emph{Upper panels}: Best-fit (red stars) and reported (\citealp[squares for][]{2023MNRAS.519.1526P}; \citealp[circles for][]{2022A&A...661L...7D}) values for the bending mass and maximum SFR, defined as $\mathcal{M}_0 \equiv \log_{10} (M_0/M_\odot)$ and $\mathcal{SFR}_{\rm max} \equiv \log_{10} (\mathrm{SFR}_{\rm max}/M_\odot\,\mathrm{yr}^{-1})$, as a function of redshift.}
    \label{fig:MSfit_full}
\end{figure}

When fitting the \gls{ms}, we must take into account the fact that for each redshift bin there are two possible values for the mass completeness limit, depending on whether consider the mass at the lower limit of the redshift bin, $z_{\rm start}$, or at the higher limit, $z_{\rm end}$. Depending on the width of the redshift bin, these two masses can differ significantly, of the order of $0.4$ -- $0.6\,\mathrm{dex}$ (see Fig.\,\ref{fig:MS_full}). When $M_\ast > M_{\rm lim} (z_{\rm end})$, we are dealing with a complete sample; in this case, we fit the points measured as the median stellar masses and \gls{sfr} of the distribution of \gls{sfg} in each bin of mass, with the associated uncertainty as the quadrature sum of the standard deviation of the median in each bin and the typical uncertainty on the \gls{sfr} (see Sect.\,\ref{sc:Validation}). These are shown as stars in the bottom panel of Fig.\,\ref{fig:MSfit_full}, coloured as a function of the redshift bin to which they belong. When $M_\ast < M_{\rm lim} (z_{\rm start})$ we consider the sample to be incomplete, and these galaxies are excluded from the fit. In the mass bins where $M_{\rm lim} (z_{\rm start}) < M_\ast < M_{\rm lim} (z_{\rm end})$, we are preferentially missing lower-mass galaxies, which tend to have lower SFRs. As a result, we treat the SFR data points in these bins as upper limits, as is indicated by the arrows in the bottom panel of Fig.\,\ref{fig:MSfit_full}.
\begin{figure}
    \centering
    \includegraphics[width=\hsize]{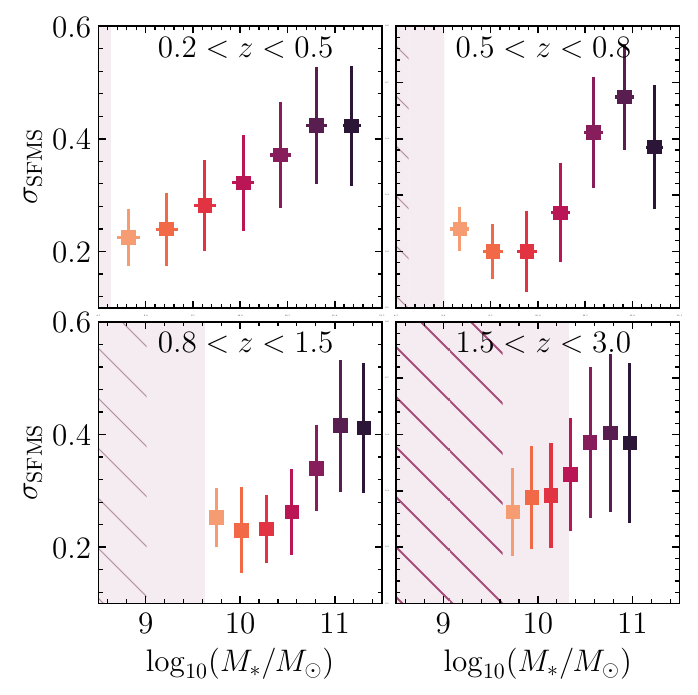}
    \caption{Scatter of the $\mathrm{SFR}$--$M_\ast$ relation, $\sigma_{\rm SFMS}$, in different mass bins. The purple shaded regions highlight where the mass is below the completeness limit at the start (hatched) and end of the redshift bin.}
    \label{fig:MS_scatter}
\end{figure}
\begin{table}[]
    \centering
    \caption{Fitted values for $\log_{10}(M_0/M_\odot)$, $\log_{10}(\mathrm{SFR}_\mathrm{max}/M_{\odot}\,\mathrm{yr}^{-1})$, and the relation scatter $\sigma_{\rm SFMS}$ for each redshift bin, with the median redshift in bin reported in the first column.}\label{tab:M0SFR0}
    \begin{tabular}{cccc}
        \hline
        \hline
        \noalign{\vskip 3pt}
        $z_{\rm med}$ &  $\log_{10} (M_0/M_{\odot})$ & $\log_{10} (\mathrm{SFR}_\mathrm{max}/M_{\odot}\,\mathrm{yr}^{-1})$ & $\sigma_{\rm SFMS}$ \\
        \hline
        \noalign{\vskip 3pt}
        $0.37$ & $10.27 \pm 0.04$ & $1.06 \pm 0.03$ & $0.34 \pm 0.11$ \\
        $0.68$ & $10.29 \pm 0.05$ & $1.29 \pm 0.03$ & $0.27 \pm 0.09$ \\
        $1.15$ & $10.62 \pm 0.05$ & $1.74 \pm 0.04$ & $0.28 \pm 0.08$ \\
        $1.83$ & $11.10 \pm 0.09$ & $2.28 \pm 0.07$ & $0.40 \pm 0.12$ \\
        \hline
    \end{tabular}
\end{table}
\begin{figure*}
    \centering
    \includegraphics[width=\hsize]{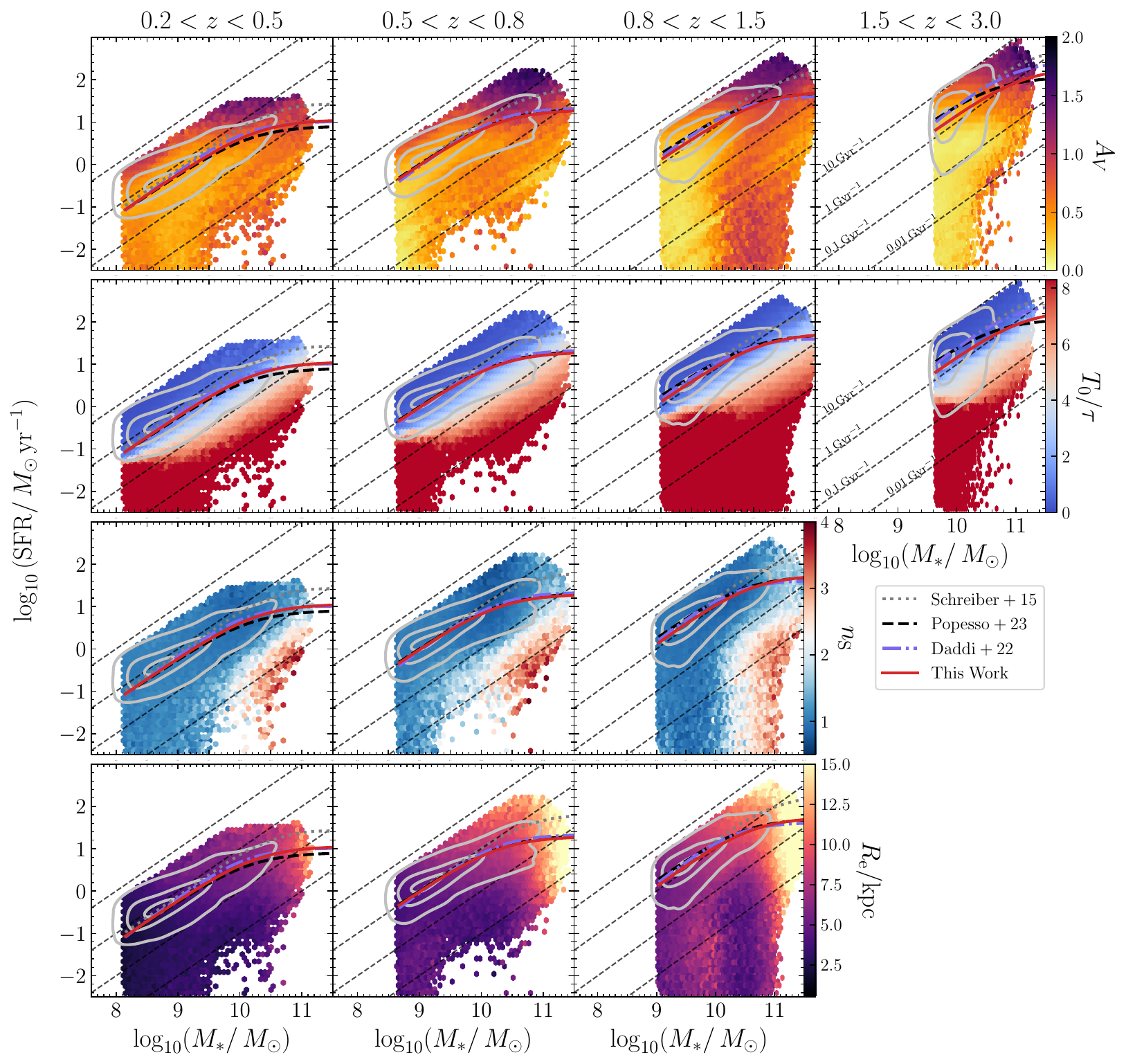}
    \caption{$\mathrm{SFR}$--$M_\ast$ relation, colour-coded by the median $A_V$ (\emph{top}), $T_0/\tau$ (\emph{middle-top}), Sérsic index (\emph{middle-bottom}), and Sérsic radius (\emph{bottom}) of objects in each bin. A bin is coloured only when the number of sources falling within the bin is higher than $20$. The contours in grey contain $90\%$, $50\%$, and $10\%$ of the full sample. The literature \gls{ms} are the same as the ones shown in Fig.\,\ref{fig:MS_full} and reported in Sect.\,\ref{sc:Results}. Dashed lines highlight the loci of equal sSFR.}
    \label{fig:MS_dep_full}
\end{figure*}

The SFR--mass bins were obtained by binning the distribution of stellar mass to uniformly cover the stellar masses space with a similar number of sources to make these statistically significant. The bins start from where the stellar mass is higher than the $95\%$ completeness limit at the lower limit of the redshift bin; that is, $\Mstarwun \sim 8.1$ at $z = 0.37$, $\Mstarwun \sim 8.6$ at $z = 0.67$, $\Mstarwun \sim 9.0$ at $z = 1.15$, and $\Mstarwun \sim 9.5$ at $z = 1.83$.  The results of the fit are reported in Table\;\ref{tab:M0SFR0}, and shown in Fig.\,\ref{fig:MSfit_full} as dashed lines colour-coded by redshift bin. The bending mass and maximum SFR are reported as a function of redshift in the upper panels of Fig.\,\ref{fig:MSfit_full}, as red stars and solid lines. We also show, for comparison, the same parameters found in \citet[][as blue circles and dashed dotted line]{2022A&A...661L...7D} and \citet[][black edged squares and dashed line]{2023MNRAS.519.1526P} as a function of $z$. As an internal check, we also fitted the same points with a linear relation \citep[e.g., Eq.\;10 in][]{2023MNRAS.519.1526P}, obtaining worse $\chi^2$ for each redshift bin with respect to the relation including the bending at the high-mass end.

Our recovered bending masses $M_0$ increase with redshift. We find results similar to those of \citet{2022A&A...661L...7D}, with the main difference in the $[0.8$, $1.5]$ bin, with a higher $\log_{10} (M_0/M_\odot) = 10.62 \pm 0.05$ \citep[similar to what was found in][]{2023MNRAS.519.1526P}. In the $[1.5$, $3.0]$ bin, our bending mass is more in line with \citet{2022A&A...661L...7D} at $\log_{10} (M_0/M_\odot) = 11.10 \pm 0.09$, \citep[$0.6\,\mathrm{dex}$ higher than the one in][]{2023MNRAS.519.1526P}. However, for this redshift bin, we must again warn the reader about the caveats highlighted in Sect.\,\ref{sc:Validation} and Fig.\,\ref{fig:MS_vs_deltaz}; that is, at $11 < \Mstarwun < 11.5$ any wrong redshift attribution will introduce an important source of systematic biases, skewing the fitted bins at higher values of $M_\ast$ and SFR, and therefore the recovered $M_0$ and $\mathrm{SFR}_{\rm max}$. In any case, the observed agreement and evolution with redshift are remarkable, when taking into account the fact that our \gls{sfr} have been evaluated with photometry ranging from the $g$ band (the $u$ band is available only in one field) up to $4.5\,\micron$, and thus without properly accounting for dust-obscured star formation processes, which are disentangled from quiescence only when photometry from the far-IR up to radio is available. 

Trying to find the best-fit parameters of the \gls{ms} with the functional form in Eq.\;(\ref{eq:MSform}), we also measured the scatter in the relation $\sigma_{\rm SFMS}$, fitting the difference between the observed set of galaxy stellar masses and \gls{sfr} and the model as a normal distribution, $\mathcal{N} (\sigma_{\rm SFMS})$. We find the same tight scatter, $\sigma_{\rm SFMS} \simeq 0.3$, observed in previous studies \citep[e.g.][for a SED-fitting derived \gls{ms}]{2024ApJ...977..133C}, consistent in all redshift intervals within the uncertainties (Table\,\ref{tab:M0SFR0}). We also investigated whether there is any correlation between $\sigma_{\rm SFMS}$ and stellar masses. The topic is still highly debated, from both theoretical and observational points of view. Simulations find either a mass dependence \citep[e.g.][]{2018MNRAS.480.4842C, 2019ApJ...879...11K, 2022MNRAS.509..595L} or no dependence at all \citep[e.g.][]{2019MNRAS.485.4817D, 2023A&A...677L...4P}. The same is true for observational studies, with reported cases of no trend \citep[e.g.][]{2015A&A...575A..74S, 2015ApJ...815...98S, 2024ApJ...977..133C}, a monotonically increasing trend \citep{2019MNRAS.483.3213P, 2021MNRAS.505..947S, 2025ApJ...979..193C}, or even a decrease with stellar mass \citep{2015MNRAS.449..820W}. In each redshift interval, we measured the \gls{ms} scatter in seven different mass bins as was previously done, starting from the mass limit completeness at the end of the interval (or at the start, for $1.5 < z < 3.0$) in order to reduce any systematic effect arising from mass incompleteness. To account for the uncertainties in redshift, stellar mass, and SFR, we measured $\sigma_{\rm SFMS}$ in each mass bin ten times, each time on a sample generated perturbing the best-fit redshifts and \gls{pp} with the measured uncertainties. We then took every measure of $\sigma_{\rm SFMS}$ as the median of all iterations, with associated uncertainties as the 16th and 84th percentile of the distributions. The results are reported in Fig.\,\ref{fig:MS_scatter}. Even though caution must be taken given the various sources of uncertainties, we observe an increase in $\sigma_{\rm SFMS}$ with stellar mass, or at least an overall increasing trend.

The \gls{ms} dependence on the absorption, $A_V$, and the ratio between the age and the e-folding time, $\tau$, is shown in the first two rows of Fig.\,\ref{fig:MS_dep_full}, where the colours correspond to the median of $A_V$ and $T_0/\tau$ values of the sources within each bin, with the \gls{ms} contours superimposed in grey. In this case, we have not limited the sample to \gls{sfg} only (i.e. we have not removed the sources identified as quiescent in the $\mathrm{NUV}$--$r^{+}$--$J$ diagram), highlighting a bin when the number of sources within the bin is greater than $20$. In this way, we can explore less populated regions of the $M_\ast$--SFR plane, where passive galaxies are expected to appear. The location of galaxies in the plane below a certain limiting \gls{sSFR} (e.g. $0.01\,\mathrm{Gyr}^{-1}$) should not be considered as absolute, but rather as indicative that those objects are passive galaxies and could lie anywhere below that line. As was expected, maximum extinction values ($A_V > 2.0$) are associated with the most massive and star-forming galaxies of the \gls{ms}, with $1 < \mathrm{sSFR}/\mathrm{Gyr}^{-1} < 10$, while low or zero dust extinction values are observed at lower sSFRs. The typical $A_V$ of \gls{ms} galaxies decreases slightly from $A_V \sim 1.0$ to $A_V \sim 0.5$ with increasing redshift. Looking at the distribution of fitted $A_V$, we recover only a small fraction of objects ($\sim 1\%$) with a high extinction of $A_V > 2.5$, with the $A_V$ distribution peaking at $0.5 < A_V < 1.0$. This could be due to the particular limitations and caveats of the recovered sample (see Sect.\,\ref{sc:Methodology}), especially in the range of wavelengths covered. Lower \gls{sSFR} galaxies are associated with stellar populations with older ages, in particular at low redshift, and high age$/\tau$ ratios, while younger ages are found in the upper part of the \gls{ms} and at high redshift \citep{2025A&A...695A..86N}.

The morphological parameters of all the \Euclid-detected sources were measured by running the \texttt{SourceXtractor++} code \citep{2020ASPC..527..461B, 2022ASPC..532..329K}, fitting the detections as two-dimensional Sérsic profiles (\citealp[see][]{Q1-TP004} \citealp[and][for further details]{Q1-SP040}). Here, we focus on the Sérsic radius, $R_e$, and the Sérsic index, $n_{\rm S}$. In order to be more conservative, following what was found in \citet[][see their Fig.\,2]{Q1-SP040}, in addition to the cuts described in Sect.\,\ref{sc:Data}, we also removed the sources whose fit did not converge on a solution. In these cases, the Sérsic axis ratio is exactly equal to one, and the Sérsic index $n_{\rm S} < 0.302$ or $n_{\rm S} > 5.45$. These cuts reduce our sample for morphological analysis by $17\%$, leaving a total of \num{6702811} sources.

In the two bottom rows of Fig.\,\ref{fig:MS_dep_full}, we report the \gls{ms}, colour-coded for the median values of $n_{\rm S}$ (middle-bottom row) and $R_{\rm e}$  (bottom row), converted from arcseconds to kiloparsecs for the predicted photo-$z$). We note that the higher values of $n_{\rm S} > 3$ are associated with the most massive, low-\gls{sSFR} objects, and the radii, $R_{\rm e}$, increase with stellar mass at all redshifts. The combination of the limited volume of Universe sampled at low $z$ with the increasing difficulty of properly identifying smaller galaxies at higher redshifts due to the fixed \Euclid resolution (roughly translating into a limit in measured effective radiuses, $R_{\rm e, lim } \simeq 2$ -- $3\,{\rm kpc}$, in the redshift interval $[0.5$, $3.0]$) results in an apparent decrease in $R_{\rm e}$ with cosmic time, as the distributions of galaxies falling within each stellar mass--SFR bin suffer from this size incompleteness.

In all redshift intervals, the concurrent presence of the \gls{ms} and the prominence of exponential disks with $n_{\rm S} \simeq 1$ (in blue) is immediately visible, while at lower \gls{sSFR}, close to the limit where measuring SFRs becomes difficult, we note a crowding of sources with de Vaucoulers profiles of $n_{\rm S} \simeq 4$ (in red). Between those, there is an intermediate region of profiles with $2 < n_{\rm S} < 3$ (the white strip) at about $\mathrm{sSFR} \sim 0.05\,\mathrm{Gyr}^{-1}$, which indicates an evolutionary trend in the structural parameters while moving in the stellar mass--SFR plane, in agreement with what has been observed previously in other samples \citep{2003MNRAS.341...54K, 2011ApJ...742...96W, 2015ApJ...811L..12W, 2019MNRAS.482L.129M, 2023ApJ...942...49C, 2025A&A...694A..76M}. This kind of transition is observed only at the medium-to-high-mass end $\Mstarwun > 10$, and appears to show only a weak redshift dependence. Interestingly, the mass over which the sequence of passive $n_{\rm S} \simeq 4$ profiles is observed is almost coincident with the fitted bending mass, $M_0$, a trend that has already been observed in the literature \citep[e.g.][]{2020ApJ...899...58L}. 

\Euclid combines a large sample area with a uniform multi-wavelength view, which has great potential for environmental studies. Despite that, here we do not focus on how the environment shapes \gls{ms}, as this is the main scope of other Q1 works \citep{Q1-SP017, Q1-SP022}. The soon-to-be-available DR1 observations of the EDFs are expected to have at least ten ROSs, which corresponds to $1.25$ magnitude-deeper data. Combining the DR1 data, along with other multi-wavelength observations and the \Euclid spectroscopy sample, will allow for a thorough investigation of the environmental effects on the \gls{ms}. 

\section{\label{sc:Conclusions}Conclusions}
The \Euclid Q1 data release, with its first look at the EDFs, is already a good test of the capabilities of the mission to investigate the formation and evolution of galaxies, especially in terms of the enhanced statistics of such a large area of the extragalactic sky. In this work, we have investigated the relation between stellar masses and \gls{sfr}, the \gls{ms}, and how it relates to the other well-constrained \gls{pp} and morphological parameters, comparing the results with the existing literature, as a fair validation of the \Euclid results and a first exploration of the mission capabilities to return a statistically robust sample. Even a single ROS of the EDFs is able to yield reliable measurements for the photometric redshifts, stellar masses, and \gls{sfr} -- which in this work have been improved due to the addition of the first two channels of IRAC -- for more than eight million galaxies at a magnitude cut of $\HE < 24$. In particular, more than $\sim 30\,\mathrm{k}$ galaxies are found with $\Mstarwun > 11$, a substantial improvement over all the other extragalactic surveys limited to a few square degrees of the sky.

The recovered scaling relation between stellar mass and SFR is consistent with what is known from previous studies on the \gls{ms}. The data show a similar tight dispersion, between $0.26$ and $0.40\,\mathrm{dex}$, with an increase in $\sigma_{\rm SFMS}$ occurring at higher masses, and are in excellent agreement with the functional forms reported in \citet{2023MNRAS.519.1526P}, \citet{2022A&A...661L...7D}, and \citet{2015A&A...575A..74S}. These works fit \gls{ms} with a non-linear term that translates into a bending of the relation, becoming more pronounced at high masses. We fit the relation with the same parameterisation -- that is, linking the SFR directly to the bending mass, $M_0$, and the SFR maximum, $\mathrm{SFR}_\mathrm{max}$ -- recovering the same statistical agreement with the previous studies. $M_0$ increases almost monotonically with redshift, starting from $\log_{10} (M_0/M_\ast) \simeq 10.3$ at $z = 0.37$, up to $11.1$ at cosmic noon. Our results are in accord with the presence of this reduction in SFR in massive galaxies, with a better $\chi^2$ with respect to a linear one. At the same time, when correlating our results with the catalogue of morphological properties, we recover the well-known bimodality between exponential and smaller star-forming disks and de Vaucoulers profiles for passive and bigger galaxies.

\begin{acknowledgements}
AnEn, LuPo, MiBo, EmDa, ViAl, FaGe, SaQu, MaTa, MaSc, and BeGr acknowledge support from the ELSA project. "ELSA: Euclid Legacy Science Advanced analysis tools" (Grant Agreement no. 101135203) is funded by the European Union. Views and opinions expressed are however those of the author(s) only and do not necessarily reflect those of the European Union or Innovate UK. Neither the European Union nor the granting authority can be held responsible for them. UK participation is funded through the UK HORIZON guarantee scheme under Innovate UK grant 10093177.
AnEn, MiBo, StCa, ElZu acknowledge support from the INAF MiniGrant 2023 "ADIEU: Anomaly Detections In
EUclid".
This work has been partially funded by Italian MUR Premiale MITiC 2017.
\AckEC
\AckQone
\AckCosmoHub
Based on data from UNIONS, a scientific collaboration using three Hawaii-based telescopes: CFHT, Pan-STARRS, and Subaru \url{www.skysurvey.cc}\,. Based on data from the Dark Energy Camera (DECam) on the Blanco 4-m Telescope at CTIO in Chile \url{https://www.darkenergysurvey.org}\,.

In preparation for this work, we used the following codes for Python: \texttt{Numpy} \citep{2020Natur.585..357H}, \texttt{Scipy} \citep{2020NatMe..17..261V}, \texttt{Pandas} \citep{mckinney-proc-scipy-2010}, \nnpz\ \citep{2018PASJ...70S...9T},

\end{acknowledgements}

\bibliography{biblio}

\label{LastPage}
\end{document}